\documentclass[aps,pre,twocolumn,showpacs,amsmath,graphicx]{revtex4}

\usepackage{array}

\usepackage{graphicx}
\usepackage{multirow}
\usepackage{subfigure}
\providecommand{\tabularnewline}{\\}
\newcommand{\lyxdot}{.}

\newcommand{\eq}{Eq.}
\newcommand{\sect}{Sec.}
\newcommand{\el}{Euler-Lagrange}
\newcommand{\chc}{CHC}
\newcommand{\fig}{Fig.}
\newcommand{\dis}{dissipative}
\newcommand{\psif}{\psi}
\newcommand{\psifh}{\hat{\psi}}
\newcommand{\phif}{\phi}

\begin{document}

\preprint{This line only printed with preprint option}

\title{Early time kinetics of systems with spatial symmetry breaking}

\author{Rachele Dominguez}
\email{erdomi@bu.edu}
\affiliation{Department of Physics, Boston University, Boston, Massachusetts 02215}

\author{Kipton Barros}
\affiliation{Department of Physics and Center for Computational Science, Boston
University, Boston, Massachusetts 02115}

\author{W. Klein}
\affiliation{Department of Physics and Center for Computational Science,  Boston
University, Boston, Massachusetts 02215}

\begin{abstract}
In this paper we present a study of the early stages of unstable state evolution of systems with spatial symmetry changes.  In contrast to the early time linear theory of unstable evolution
described by Cahn, Hilliard, and Cook, we develop a generalized theory
that predicts two distinct stages of the early evolution for symmetry breaking
phase transitions. In the first stage the dynamics is dominated by
symmetry preserving evolution. In the second stage, which
shares some characteristics with the Cahn-Hilliard-Cook theory, noise
driven fluctuations break the symmetry of the initial phase on a time
scale which is large compared to the first stage for systems with long interaction
ranges. To test the theory we present the results of numerical simulations
of the initial evolution of a long-range antiferromagnetic Ising model
quenched into an unstable region. We investigate two types of symmetry
breaking transitions in this system: disorder-to-order and order-to-order
transitions. For the order-to-order case, the Fourier modes evolve as a linear combination of exponentially growing or decaying terms with different time scales.
\end{abstract}
\maketitle

\section{introduction\label{sec:introduction}}

Many systems have effective long-range interactions and exhibit a
variety of crystalline phases. Some examples are metals~\citep{Christian2002}, block copolymers~\citep{Bates1990}, 
and systems with modulated phases~\citep{Seul1995,Garel1982,Lacoste2001}. The kinetics of
the transitions from a disordered, spatially uniform state to a spatially
periodic state (disorder-to-order) have been explored in several contexts
(see, for example, Refs.~\citep{Harrison2000,Roland1990,Wickham2003,Fredrickson1989}.)
Some aspects of the kinetics from one spatially periodic state to
another (order-to-order) have also been investigated~\citep{Li2007a,Li2007,Yu2005,Nonomura2001}.

For systems with long-range interactions the phase ordering
process is separated into an early time regime described by a linear theory,
an intermediate regime where nonlinear effects first become important,
and a late stage domain coarsening regime. Phase ordering, a process whereby a disordered state evolves into a state with macroscopic regions of like components,  has been well studied for systems described
by a Landau-Ginzburg free energy, such as the ferromagnetic Ising model~\citep{Gunton1983}. If the coarse-grained order
parameter is locally conserved, the phase ordering process
is called spinodal decomposition. If it is not conserved, the process
is called continuous ordering. The latter case is referred to as Model
A and the former is referred to as Model B~\citep{Hohenberg1977}. 

``Phase ordering'' as used here is distinct from a disorder-to-order
transition, which we take to mean a transition
from a disordered to a spatially periodic state.  A disorder-to-order transition breaks spatial symmetry; that is, the initial state has a spatial
symmetry that is absent in the final state. In contrast,
phase ordering for the ferromagnetic Ising model with a non-conserved order parameter has uniform spatial symmetry in both the initial and final phases. Thus, in this case phase ordering does not break spatial symmetry~\cite{symmetry}.

The early time linear regime of phase ordering
was first studied by Cahn and Hilliard~\citep{Cahn1958,Cahn1959,Cahn1967}
and later by Cook~\citep{Cook1970}. The early time theory was extended
to the intermediate time regime by Langer, Bar-on, and Miller~\citep{Langer1975}
for Model B and by Billotet and Binder~\citep{Billotet1979} for
Model A. The Lifshitz-Slyozov theory~\citep{Lifshitz1961} describes
the late time regime for Model B which is dominated by the surface
tension of domain walls. The theory of late stage boundary motion
for model A was developed by Allen and Cahn~\citep{ALLEN1979}.

The Cahn, Hilliard, and Cook (\chc) linear theory was originally developed to describe spinodal decomposition for systems such
as binary alloys, where the order parameter is locally conserved. The \chc\ theory predicts exponential growth of certain Fourier modes immediately
after a temperature quench into the unstable regime~\citep{Cahn1958,Cahn1959,Cahn1967,Cook1970}.
For systems with long-range interactions, this exponential growth
lasts for a time proportional to the logarithm of the interaction range~\citep{Binder1984}.

\chc\ theory also describes continuous ordering for systems such
as the ferromagnetic Ising model, with a non-conserved order parameter. In this case the temperature is quenched to
below the critical temperature at zero external field (a critical
quench). However, if the external field is non-zero during the temperature
quench (an off-critical quench), \chc\ theory is not applicable. We will
show that \chc\ theory also fails to describe a wide class of transitions
that break spatial symmetry.

In this paper we extend the \chc\ theory to describe transitions that
break spatial symmetry and whose dynamics do not conserve the order parameter~\cite{analogy}. For these systems we find two distinct stages of the early
time evolution in contrast to \chc. Stage one, which is absent in \chc\ theory, is dominated
by symmetry preserving dynamics. The time scale for which this stage applies is independent of the interaction range. Stage two is analogous to the early time evolution
described by \chc\ because the dynamical equations are characterized by exponential solutions,
and the time scale for which stage two is valid is longer for longer interaction
ranges. For disorder-to-order transitions certain Fourier modes grow exponentially. In contrast, for order-to-order transitions the Fourier modes evolve as a linear combination of exponentially growing or decaying terms with different time scales.

Our particular system of interest is the long-range antiferromagnetic
Ising model. This system has spatial symmetry
breaking disorder-to-order and order-to-order transitions.
We use Monte Carlo techniques to simulate the system and compare our
results to our numerical solutions of a Langevin equation and to
our analytical predictions. 

The structure of the paper is as follows. Section~\ref{sec:background-on-unstable}
summarizes the theoretical background on unstable state evolution for models
described by a Landau-Ginzburg free energy, focusing on the linear
\chc\ theory. We extend this
theory to include symmetry breaking transitions in \sect~\ref{sec:The-onset-of}. The role that symmetry breaking
plays in the interplay between the two stages is discussed. The long-range
antiferromagnetic Ising model and its phase diagram are introduced
in \sect~\ref{sec:the-anti-ferromagnetic-long}. We develop the
generalized theory for this model by using the Langevin equation
based on a coarse-grained free energy as well as the Langevin equation
describing Glauber dynamics. Our numerical and theoretical results
are presented for the disorder-to-order case in \sect~\ref{sec:disorder-to-order}
and for the order-to-order case in \sect~\ref{sec:order-to-order}.
Section~\ref{sec:summary-and-discussion} summarizes and
discusses our results.

\section{review of CHC theory for continuous ordering\label{sec:background-on-unstable}}

In this section we review the Cahn-Hilliard-Cook (\chc) theory as it is applied to continuous ordering for which the order parameter is not conserved. We begin with the Langevin equation for the evolution of the order
parameter $\hat{\phi}(\boldsymbol{x},t)$:
\begin{equation}
\frac{\partial\hat{\phi}(\boldsymbol{x},t)}{\partial t}=-M\left(\frac{\delta F[\hat{\phi}]}{\delta\hat{\phi}(\boldsymbol{x})}\right)+\sqrt{B}\eta(\boldsymbol{x},t),\label{eq:generalLangevin1}
\end{equation}
where $F$ is the coarse grained free energy, and $M$ is the mobility
(which for the \chc\ theory is assumed to be a constant in space and time) and
\begin{equation}
B=2MT\label{eq:B}
\end{equation}
from the fluctuation dissipation relationship where we have set $k_B=1$. Equation~\eqref{eq:generalLangevin1}
is also known as the time-dependent Ginzburg-Landau equation~\citep{Bray1995}. The first term on the right-hand side of \eq~\eqref{eq:generalLangevin1},
which we call the drift term, lowers the free energy as time evolves. The second term
describes the noise and represents the small scale fast dynamics as
random forces. We will take the noise to be white Gaussian and to obey
the conditions:
\begin{align}
\left\langle \eta(\boldsymbol{x},t)\right\rangle  & =  0\label{eq:noiseAverage}\\
\left\langle \eta(\boldsymbol{x},t)\eta(\boldsymbol{x'},t')\right\rangle  & =  \delta(\boldsymbol{x}-\boldsymbol{x'})\delta(t-t').\label{eq:noiseConditions}
\end{align}
By using the the Landau-Ginzburg free energy,
\begin{align}
F[\hat{\phi}(\boldsymbol{x})] & =  \!\int\! d\boldsymbol{x}\Big(\frac{R^{2}}{2}[\nabla\hat{\phi}(\boldsymbol{x})]^{2} +\frac{\epsilon}{2}\hat{\phi}^{2}(\boldsymbol{x}) \nonumber \\
&\quad  +\frac{1}{4}\hat{\phi}^{4}(\boldsymbol{x})-h\hat{\phi}(\boldsymbol{x})\Big),\label{eq:landau-ginsberg}\end{align}
 \eq~\eqref{eq:generalLangevin1} becomes,
 \begin{eqnarray}
\frac{\partial\hat{\phi}(\boldsymbol{x},t)}{\partial t} & = & -M\{-R^{2}\nabla^{2}\hat{\phi}(\boldsymbol{x},t)+\epsilon\hat{\phi}(\boldsymbol{x},t)\nonumber \\
 &&{} +\hat{\phi}^{3}(\boldsymbol{x},t)-h\}+\sqrt{B}\eta(\boldsymbol{x},t),\label{eq:langevin}
 \end{eqnarray}
where $\epsilon=(T-T_c)/T_{c}$ is the reduced temperature,
$T_{c}$ is the critical temperature, $h$ is the external field, and $R$ is the interaction range. The extrema of the free energy can be found by
solving the \el\ equation:
\begin{align}
0 & =  \frac{\delta F[\hat{\phi}]}{\delta\hat{\phi}(\boldsymbol{x})}\nonumber\\
& =  -R^{2}\nabla^{2}\hat{\phi}(\boldsymbol{x},t)+\epsilon\hat{\phi}(\boldsymbol{x},t)+\hat{\phi}^{3}(\boldsymbol{x},t)-h.\label{eq:eulerLagrange}
\end{align}

For a critical quench the initial state is equilibrated at initial temperature $T_i>T_c$
in zero external field $h_{i}=0$. The temperature
is instantaneously quenched to a final temperature $T_{f}<T_{c}$ in the unstable regime
where $\epsilon_{f}=(T_{f}-T_{c})/T_{c}<0$, keeping the external
field $h_{f}=0$. Before the quench, the system is in a
disordered state corresponding to the average order parameter, $\hat{\phi}_{0}=\langle\hat{\phi}(\boldsymbol{x})\rangle=0$
for spatially homogeneous systems. We will consider small fluctuations,
$\hat{\psi}(\boldsymbol{x},t)$, of the order parameter about this value,
$\hat{\phi}(\boldsymbol{x},t)=\hat{\phi}_{0}+\hat{\psi}(\boldsymbol{x},t)=\hat{\psi}(\boldsymbol{x},t),$
and keep terms first order in $\hat{\psi}(\boldsymbol{x},t)$ so that
\eq~\eqref{eq:langevin} becomes:
\begin{align}
\frac{\partial\hat{\psi}(\boldsymbol{x},t)}{\partial t} & =  -M\left(-R^{2}\nabla^{2}\hat{\psi}(\boldsymbol{x},t)-|\epsilon_{f}|\hat{\psi}(\boldsymbol{x},t)\right)\nonumber \\ & \qquad +\sqrt{B}\eta(\boldsymbol{x},t).\label{eq:linear}\end{align}
We have used the fact that $\hat{\phi}_{0}=0$ is a solution to \eq~\eqref{eq:eulerLagrange}.
In Fourier space \eq~\eqref{eq:linear} becomes \begin{equation}
\frac{\partial\psifh(\boldsymbol{k},t)}{\partial t}=D_{\rm chc}(k)\psifh(\boldsymbol{k},t)+\sqrt{B}\eta(\boldsymbol{k},t),\label{eq:linearK}
\end{equation}
with
\begin{equation}
D_{\rm chc}(k)=-M(R^{2}k^{2}-|\epsilon_{f}|).\label{eq:D_k}\end{equation}
We see from \eq~\eqref{eq:linearK} that the individual Fourier modes
evolve independently of one another. The solution to \eq~\eqref{eq:linearK}
is
\begin{eqnarray}
\psifh(\boldsymbol{k},t) & = & \psifh(\boldsymbol{k},0)\exp\left(D_{\rm chc}(k)t\right) \nonumber \\
&&{} +\sqrt{B}\!\int_{0}^{t}\eta(\boldsymbol{k},t)G(t-t')dt',\label{eq:linearSolK}
\end{eqnarray}
where $G(t-t')=\exp((t-t')D_{\rm chc}(k))$. We
use the definition of the equal time structure factor $S(\boldsymbol{k},t)=\left\langle \phif(\boldsymbol{k},t)\phif(-\boldsymbol{k},t)\right\rangle/V,$ \eq~\eqref{eq:B},
and the properties of the noise (Eqs.~\eqref{eq:noiseAverage} and~\eqref{eq:noiseConditions}) to obtain:
\begin{eqnarray}
S(k,t) & = & e^{2D_{\rm chc}(k)t}\left(S(k,t=0)+\frac{M}{\beta_{f}D_{\rm chc}(k)}\right)\nonumber \\
&&{} -\frac{M}{\beta_{f}D_{\rm chc}(k)},\label{eq:struct_factor}
\end{eqnarray}
where $\beta_{f}=T_{f}^{-1}$.

Systems with $R\gg1$
are said to be near-mean-field and the limit $R\rightarrow\infty$
is the mean-field limit~\citep{klein2007}. For $T_f<T_c$ simulations have shown that \eq~\eqref{eq:struct_factor} is applicable for systems with $R\gg1$ for some initial time after the quench~\citep{Heermann1984,Heermann1985,Izumitani1985,Okada1986}.
Binder~\citep{Binder1984} has argued that the linear theory should break down when the fluctuations are
large in comparison to the average order parameter. He predicted that \chc\ theory breaks down at a time $\tau$ that scales as
$\tau\sim\ln{R}$ (see \eq~\eqref{eq:binder result}).

To understand Binder's result, we consider how $\psifh(\boldsymbol{k})$
depends explicitly on $R$, the relevant length scale of the problem. We scale all lengths with
$R$ by defining $\boldsymbol{r\equiv\boldsymbol{x}}/R,$
and $\boldsymbol{q}\equiv\boldsymbol{k}R.$
We see from \eq~\eqref{eq:D_k} that $D_{\rm chc}(q)$ is independent of $R$. It follows that $G(t-t')$ is also independent
of $R$, and from \eq~\eqref{eq:linearSolK} $\psifh(\boldsymbol{k},t)$, $\psifh(\boldsymbol{k},0)$, and $\eta(\boldsymbol{k},t)$  must depend explicitly on $R$ in the same way. Equations~\eqref{eq:noiseAverage} and~\eqref{eq:noiseConditions} become
\begin{align}
\left\langle \eta(\boldsymbol{r},t)\right\rangle  & =  0\label{eq:scaled_noise_average}\\
\left\langle \eta(\boldsymbol{r},t)\eta(\boldsymbol{r'},t')\right\rangle  & =  \delta(\boldsymbol{r}-\boldsymbol{r'})\delta(t-t').\label{eq:scaled_noise_correlation}
\end{align}
From \eq~\eqref{eq:scaled_noise_correlation} and $\delta(\boldsymbol{x})= R^{-d}\delta(\boldsymbol{r})$ 
the noise must satisfy $\eta(\boldsymbol{x},t)= R^{-d/2}\eta(\boldsymbol{r},t)$,
or
\begin{equation}
\eta(\boldsymbol{k},t)= R^{-d/2}\eta(\boldsymbol{q},t),\label{eq:noise_R_scaling}
\end{equation}
where $\eta(\boldsymbol{q},t)=\int d\boldsymbol{x}\,\rm{e}^{-i\boldsymbol{k}\cdot\boldsymbol{x}}\,\eta(\boldsymbol{r},t).$ This motivates us to introduce $\psi(\boldsymbol{q},t)$ through
\begin{equation}
\psifh(\boldsymbol{k},t)= R^{-d/2}\psi(\boldsymbol{q},t).\label{eq:psi_scaling}
\end{equation}
and $\psi(\boldsymbol{r},t)$ though $\psi(\boldsymbol{q},t)=\int d\boldsymbol{x}\,\rm{e}^{-i\boldsymbol{k}\cdot\boldsymbol{x}}\,\psi(\boldsymbol{r},t)$ and is independent of $R$. We consider the dynamics of $\psi(\boldsymbol{r}).$  If we substitute Eqs.~\eqref{eq:noise_R_scaling} and~\eqref{eq:psi_scaling} into \eq~\eqref{eq:linear} and use a Laplacian with respect to scaled coordinates $\nabla^2_r=R^2\nabla^2$ all $R$ dependence in \eq~\eqref{eq:linear} cancels. Thus the solution to \eq~\eqref{eq:linear}, $\psi(\boldsymbol{r},t)$, is also independent of $R.$ Equations~\eqref{eq:noise_R_scaling} and~\eqref{eq:psi_scaling} mean that larger $R$ results
in effectively smaller noise and noise induced fluctuations.

We may drop the $\hat{\psi}^{3}(\boldsymbol{x})$
term in \eq~\eqref{eq:linear} only if $\hat{\psi}(\boldsymbol{x})\gg\hat{\psi}^{3}(\boldsymbol{x})$. Thus we estimate the breakdown time $\tau$ from the condition $ \hat{\psi}^{2}(\boldsymbol{x})\sim1$. For a given value of $q$ this condition gives 
\begin{equation}
\frac{\exp\left(2D(q)\tau\right)}{R^{d}} \sim  1,
\end{equation}
which yields
\begin{equation}
\tau\sim\frac{d\ln R}{2D(q)}.\label{eq:binder result}\end{equation}
where we have used \eq~\eqref{eq:psi_scaling} along with the time dependence
of \eq~\eqref{eq:linearSolK}. Note that the result for the breakdown of the linear theory is a consistency
argument, and it is possible for the linear theory to fail earlier
than the time given in \eq~\eqref{eq:binder result}. 

The result that the duration of applicability of the linear theory is longer for larger
$R$ is illuminated by a simple mechanical analogy (see Fig.~\ref{fig:ballRoll}). The \chc\ theory
describes the evolution of certain fluctuations
of the system about an unstable stationary point of the free energy.
The system is analogous to a ball sitting at the top of a fairly flat hill subject to small random perturbations. These perturbations will eventually cause the ball to roll off the hill; the smaller the perturbations, the longer it takes for the ball to roll off. As $R$ increases the noise effectively decreases and the system remains near the unstable point of the free energy
longer. When the free energy can no longer be approximated quadratically, the \chc\ theory is no longer valid.

\section{Generalization of CHC theory\label{sec:The-onset-of}}

Equation~\eqref{eq:generalLangevin1}
is an example of a \dis\ Langevin equation. Because we will also consider
Glauber dynamics, we consider a Langevin equation of the more general form,
\begin{equation}
\frac{\partial \phi(\boldsymbol{r},t)}{\partial t}=\Theta[\phi,\boldsymbol{r}]+\sqrt{B[\phi, \boldsymbol{r}]}R^{-d/2}\eta(\boldsymbol{r},t),\label{eq:generalLangevin}
\end{equation}
where $\phi(\boldsymbol{r},t)$ is the order parameter with scaled coordinates, the drift term $\Theta[\phi, \boldsymbol{r}]$ is nonlinear in $\phi(\boldsymbol{r},t)$ and the noise
satisfies Eqs.~\eqref{eq:scaled_noise_average} and~\eqref{eq:scaled_noise_correlation}.
Equation~\eqref{eq:generalLangevin} is written in scaled coordinates so that the $R$-dependence of the noise term is explicit ($\Theta[\phi,\boldsymbol{r}]$ and $B[\phi,\boldsymbol{r}]$ are independent of $R$.) 

In \sect~\ref{sec:background-on-unstable} we derived the \chc\ theory as an expansion
about a uniform order parameter background. This background configuration was the solution to the \el\ equation for the parameters both before and after the quench and is therefore time independent after a critical quench. More generally, we will expand about a background configuration $\phi_b(\boldsymbol{r},t)$ which is not necessarily uniform and may be time dependent after the quench.  

At the time of the quench, the system is in a configuration consistent with the equilibrium state before the quench.  We can write this state as $\phi(\boldsymbol{r},t=0)=\phi_b(\boldsymbol{r},t=0)+R^{-d/2}\psi(\boldsymbol{r},t=0).$  The background configuration $\phi_b(\boldsymbol{r},t=0)$ is a solution to the \el\ equation for $T_i$ and $h_i$, is independent of the noise, and may be nonuniform. The term $R^{-d/2}\psi(\boldsymbol{r},t=0)$ represents small fluctuations about the background due to the noise. In general, $\phi_b(\boldsymbol{r},t=0)\neq \phi_0(\boldsymbol{r});$ that is, the background configuration is not a stationary point of the free energy and may be time dependent after the quench.  We are interested in how this background configuration and these fluctuations evolve with time after the quench, that is the time evolution of
\begin{equation}
\phi(\boldsymbol{r},t)=\phi_{b}(\boldsymbol{r},t)+R^{-d/2}\psi(\boldsymbol{r},t)\label{eq:expansion}.
\end{equation}
If we expand \eq~\eqref{eq:generalLangevin}
to first order in the small fluctuations, $R^{-d/2}\psi(\boldsymbol{r},t)$, about the background configuration $\phi_b(\boldsymbol{r},t)$, we obtain
\begin{align}
&\frac{\partial\phi_{b}(\boldsymbol{r},t)}{\partial t} + R^{-d/2}\frac{\partial\psi(\boldsymbol{r},t)}{\partial t} =
\Theta[\phi,\boldsymbol{r}]\vert_{\phi=\phi_{b}}\nonumber \\ &  \quad 
+  \!\int\! d\boldsymbol{r}'\frac{\delta\Theta[\phi,\boldsymbol{r}]}{\delta\phi(\boldsymbol{r}',t)}\Big|_{\phi=\phi_{b}}R^{-d/2}\psi(\boldsymbol{r}',t) \nonumber \\
& \quad +\sqrt{B[\phi,\boldsymbol{r}]}\Big|_{\phi=\phi_{b}}R^{-\mbox{}d/2}\eta(\boldsymbol{r},t).\label{eq:fullLangevin} 
\end{align}

It is natural to separate \eq~\eqref{eq:fullLangevin} terms of order $R^0$ from terms of order $R^{-d/2}$ giving the dynamical
equations for the background and the fluctuations. The equation for the evolution of the background
configuration, which is zeroth order in $R^{-d/2}$, is\begin{align}
\frac{\partial\phi_{b}(\boldsymbol{r},t)}{\partial t}  =  \Theta[\phi]\vert_{\phi=\phi_{b}}.   \hspace{0.2cm} \mbox{(evolution of background)} \label{eq:noiseless}\end{align}
Equation~\eqref{eq:noiseless} is noiseless and independent of $R$. In contrast, the equation for the evolution
of the fluctuations is noise driven and is first order in $R^{-d/2}$:
\begin{eqnarray}
\frac{\partial\psi(\boldsymbol{r},t)}{\partial t} & = & \!\int\! d\boldsymbol{r}'\frac{\delta\Theta[\phi,\boldsymbol{r}]}{\delta\phi(\boldsymbol{r}',t)}\Big|_{\phi=\phi_{b}}\psi(\boldsymbol{r}',t)\nonumber \\
 &&{} +\sqrt{B[\phi]}\mid_{\phi=\phi_{b}}\eta(\boldsymbol{r},t).\nonumber \\
 &&{}\qquad \mbox{(evolution of fluctuations)}\label{eq:generalLinear}
 \end{eqnarray}
 The background configuration $\phi_b(\boldsymbol{r},t)$ and the fluctuations $\psi(\boldsymbol{r},t)$ evolve simultaneously; Eqs.~\eqref{eq:noiseless} and~\eqref{eq:generalLinear}
together describe the early time kinetics of unstable systems. Equation~\eqref{eq:generalLinear}
differs from the \chc\ theory described in \sect~\ref{sec:background-on-unstable} because
the coefficient of $\psi(\boldsymbol{r},t)$ on the right-hand side
of \eq~\eqref{eq:generalLinear} depends on $\phi_{b}(\boldsymbol{r},t)$,
which evolves according to \eq~\eqref{eq:noiseless}.
Because of the explicit time dependence of $\phi_{b}(\boldsymbol{r},t)$,
we do not expect to find exponential solutions as found in the
\chc\ theory. 

Equation~\eqref{eq:noiseless} dominates the evolution of $\phi(\boldsymbol{r},t)$ until $\big{\vert}\Theta[\phi]\vert_{\phi=\phi_{b}}\big{\vert} \sim R^{-d/2}$ (we will discuss specific examples of this evolution in the following). We refer to the kinetics while the background configuration is evolving
as \textit{stage one}. If
\begin{equation}
\big{\vert}\Theta[\phi]\vert_{\phi=\phi_{b}}\big{\vert} \ll R^{-d/2},\label{eq:chc-condition}
\end{equation}
we refer to the kinetics as \textit{stage two}. 

If $\Theta[\phi]\propto\delta F/\delta\phi$, then \eq~\eqref{eq:generalLangevin} becomes a \dis\ Langevin equation. In the limit $R\rightarrow \infty$ the condition for stage two evolution (\eq~\eqref{eq:chc-condition}) implies that $\delta F/\delta\phi\simeq 0$, or\begin{equation}
\phi_{b}(\boldsymbol{r},t)=\phi_{0}(\boldsymbol{r}),\label{eq:chc-diss-condition}
\end{equation}
where $\phi_0(\boldsymbol{r})$ is an unstable solution to the \el\ equation for the final parameters $T_f$ and $h_f$ corresponding to the temperature and external field after the quench, respectively. In stage two the dynamics
shares several features with the dynamics of \chc. Namely,
the background configuration is nearly a stationary point of the free energy, and the linear evolution admits (nearly)
exponential solutions, which are applicable for a time $\tau\sim\ln R$, as in \eq~\eqref{eq:binder result}. 

If $\phi_{b}(\boldsymbol{r},t=0)\neq\phi_{0}(\boldsymbol{r})$,
then stage one kinetics begins immediately after the quench, and either there
is no stage two kinetics at any time after the quench, or stage two kinetics
follows stage one. If $\phi_{b}(\boldsymbol{r},t=0)=\phi_{0}(\boldsymbol{r})$, then stage one kinetics is absent and stage
two kinetics begins immediately after the quench. Table~\ref{tab:summary_table} summarizes the various 
possibilities. 

\begin{table*}[t]
\begin{tabular}{|l|l|l|l|l|}
\hline 
Case & Spatial symmetry breaking & Stage one & Stage two & Example\\
\hline
\multirow{2}{*}{$\phi_{b}(t=0)\simeq \phi_{0}$} & No & No & Yes & Critical quench for ferromagnet \\
\cline{2-5}
& Yes & No & Yes & Critical quench for antiferromagnet\\
\hline 
\multirow{2}{*}{$\phi_{b}(t=0)\neq\phi_{0}$} & No & Yes & No & Off-critical quench for ferromagnet\\
\cline{2-5} 
& Yes & Yes & Yes & Off-critical and external field \\
&&&& quench for antiferromagnet\\
\hline
\end{tabular}
\caption{Summary of early kinetics for various types of transitions for
the long-range ferromagnetic Ising model and the long-range antiferromagnetic
Ising model. In the table $\phi_b(t=0)$ is the background configuration just after the quench and $\phi_0$ is the unstable solution to the \el\ equation for $T_f$ and $h_i$.  These quenches are defined in the text.}
\label{tab:summary_table}
\end{table*}

\begin{figure}[h]
\subfigure[\label{fig:ballRoll.a}]{
\includegraphics[scale=0.75]{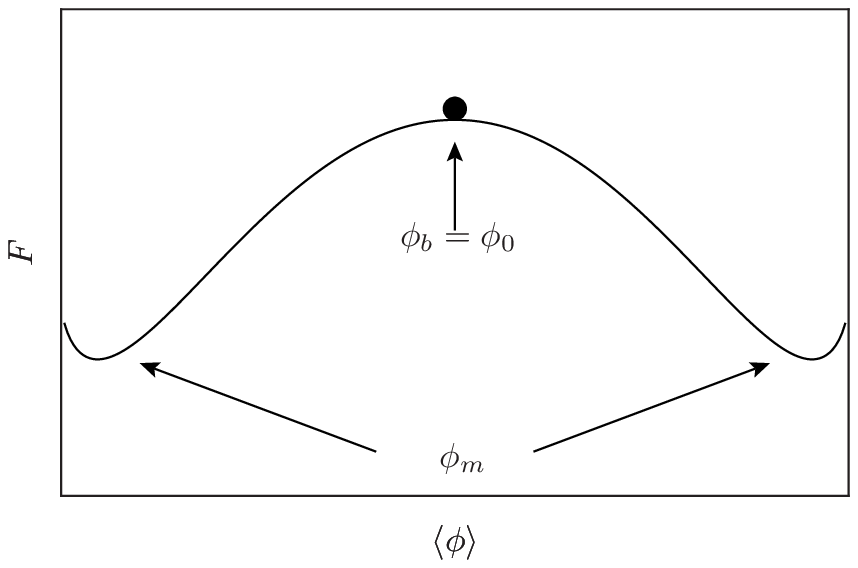}}
\quad
\subfigure[\label{fig:ballRoll.b}]{
\includegraphics[scale=0.75]{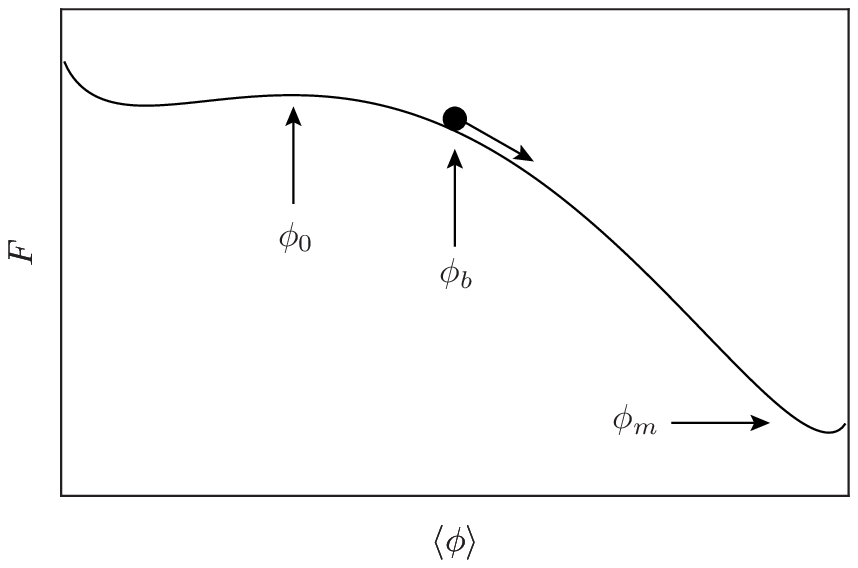}}
\caption{A representation of the free energy landscape for a ferromagnetic
Ising model as a function of the average magnetization $\langle\phi\rangle$
for (a) a critical quench and (b) an off-critical quench for $h_{f}>0.$ The dot represents the background configuration of the system immediately after the
quench. For a critical quench the system is initially at an unstable stationary
point of the free energy, and the \chc\ theory governs its early
evolution. For long-range systems the noise is effectively small, and the system fluctuates about the top of the hill for an extended time. Eventually, the system will be in one of the two global minima $\phi_m$ if the system size and temperature are finite. For an off-critical quench the background configuration
sits to the right of the free energy stationary point. The noiseless
dynamics of \eq~\eqref{eq:noiseless} drives the system directly
toward the stable minimum $\phi_m$ and away from the stationary point. Thus, in this case
we do not expect stage two exponential growth at any time after the
quench.\label{fig:ballRoll}}
\end{figure}

A critical quench in the ferromagnetic Ising model (see Table~\ref{tab:summary_table} and \fig~\ref{fig:ballRoll.a}) exhibits stage two but not stage one kinetics. In this case $T_i=\infty$, $h_{i}=0$, $T_f<T_c$, $h_f=0$, and $\phi_b(t=0)=\phi_0$. 

In contrast, an off-critical quench in the ferromagnetic Ising model (see Table~\ref{tab:summary_table} and \fig~\ref{fig:ballRoll.b}) is an example
 that exhibits stage one but not stage two kinetics. In the off-critical quench the initial conditions are the same as for the critical quench ($T_{i}=\infty$
and $h_{i}=0$), but the system is quenched below $T_c$ and
$h_f\ne0$. Because $h_f\neq0,$ the initial background
configuration $\phi_{b}(t=0)=0$ is not a solution to the \el\ equation
for the final parameters $T_f$ and $h_f$.  The background configuration evolves via \eq~\eqref{eq:noiseless}
directly toward the stable minimum $\phi_m$ of the free energy landscape and
away from the unstable solution $\phi_0$ (see \fig~\ref{fig:ballRoll.b}). Hence, the system
will never reach the unstable stationary point, and there will be no
stage two kinetics.

\begin{figure}
\includegraphics[scale=0.45]{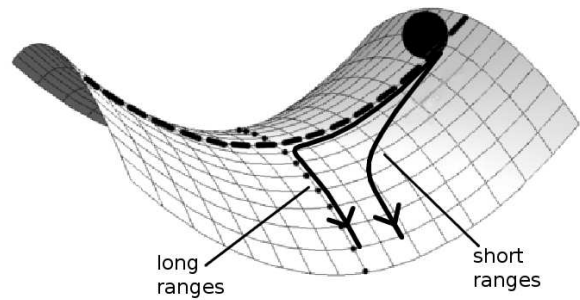}\caption{A representation of the free energy landscape for a system with spatial symmetry breaking. For systems with short range interactions fluctuations cause the ball (representing the background configuration of the system) to roll off the hill, the ball does not approach the saddle point and stage two behavior is not observed.  In contrast, for systems with long-range interactions the evolution of fluctuations governed by \eq~\eqref{eq:generalLinear} is suppressed by a factor of $R^{-d/2},$ and the noiseless dynamics of \eq~\eqref{eq:noiseless} dominates. Due to this noiseless symmetry preserving dynamics the ball rolls down the hill toward the saddle point  
along the dashed line with little deviation due to the fluctuations. The dashed line represents
all configurations with the same symmetry as the initial state. The
directions perpendicular to the dashed line represent symmetry breaking
degrees of freedom.  Once near the saddle point, the ball essentially stops rolling (condition~\eqref{eq:chc-condition} is satisfied) and stage two kinetics takes over as unstable fluctuations grow for a time of order $\tau \sim \ln R$. Eventually, after stage two, the ball will roll off the hill along the dotted line.\label{fig:saddle_point}}
\end{figure}

In contrast to the ferromagnetic Ising model~\cite{note}, we find stage one followed by stage two kinetics in systems that break spatial symmetry. In this case the noiseless dynamics in stage one and the initial
configuration conspire to drive the configuration toward the unstable
solution of the \el\ equation $\phi_0$.  In stage one both $\phi_b(\boldsymbol{r})$ and $\psi(\boldsymbol{r},t)$ evolve simultaneously 
until the system reaches the unstable solution $\phi_0$
for $T_{f}$ and $h_{f}$. Stage one kinetics is analogous to a ball rolling down a hill toward a saddle point
(see \fig~\ref{fig:saddle_point}). When the system reaches the stationary point, stage two begins. Stage two is analogous to the ball fluctuating about the top of the hill and ends when the ball finally rolls off the top of the hill towards a stable state.

Why does the ball in \fig~\ref{fig:saddle_point} not roll off
the hill before it reaches the saddle point? Symmetry considerations
provide the answer. For dissipative dynamics where
$\Theta\propto-\delta F/\delta\phi$, the noiseless dynamics of \eq~\eqref{eq:noiseless}
preserves certain spatial symmetries, depending on the form
of the free energy.  The dynamics of \eq~\eqref{eq:noiseless}
 evolves $\phi_b(\boldsymbol{r},t)$ to the lowest free energy within these symmetry constraints. This configuration is an unstable
stationary point of the free energy. It is a local minimum with respect to the symmetry
constrained degrees of freedom, but a local maximum with respect to
the other symmetry breaking degrees of freedom. Once at this saddle point, the nonlinear
background dynamics of stage one terminates, and stage two kinetics takes over. See also Ref.~\citep{Barros2008}.

The spatial symmetries preserved by the noiseless dynamics of \eq~\eqref{eq:noiseless} are not preserved by the noise driven dynamics of \eq~\eqref{eq:generalLinear}; that is, the noise causes random fluctuations which break the spatial symmetries of $\phi_b(\boldsymbol{r},t).$ Initially, these spatial symmetries are broken by $\psi(\boldsymbol{r},t=0),$ the fluctuations about the background at the time of the quench due to $\eta(\boldsymbol{r},t<0),$ the noise in the system before the quench. These initial fluctuations, which are of order $R^{-d/2},$ begin to grow (or decay) immediately after the quench according to \eq~\eqref{eq:generalLinear}. However, during stage one the fluctuations $\psi(\boldsymbol{r},t)$ remain of order $R^{-d/2}.$  The reason the ball in \fig~\ref{fig:saddle_point} does not roll off the hill is because the fluctuations which would drive the ball off the hill are small for sufficiently large $R.$ The time it takes for the ball to roll down to the saddle point (the duration of stage one), is independent of $R.$  In
stage two kinetics, when the nonlinear dynamics of \eq~\eqref{eq:noiseless} has ceased, the symmetry
breaking fluctuations continue to grow for an extended period of time, eventually becoming of the order $1.$

\section{the long-range antiferromagnetic Ising model\label{sec:the-anti-ferromagnetic-long}}

\subsection{The model and phase diagram}

To investigate a system with spatial symmetry breaking
phase transitions, we simulate the long-range antiferromagnetic Ising
model using Glauber Monte Carlo dynamics~\citep{Newman1999}. At each update a spin is selected at random and flipped with the transition probability $W=(1+\rm e^{\beta \Delta E})^{-1},$ where $\Delta E$ is the change in energy incurred from flipping the spin. Time is measured in units of Monte Carlo step per spin. For $N$ spins in a system, one Monte Carlo step (MCS) is equal to $N$ spin flip attempts. With this definition, it is natural to consider fractional Monte Carlo steps in which fewer than $N$ spin flips have been attempted.
 
\begin{figure}
\includegraphics[width=8cm]{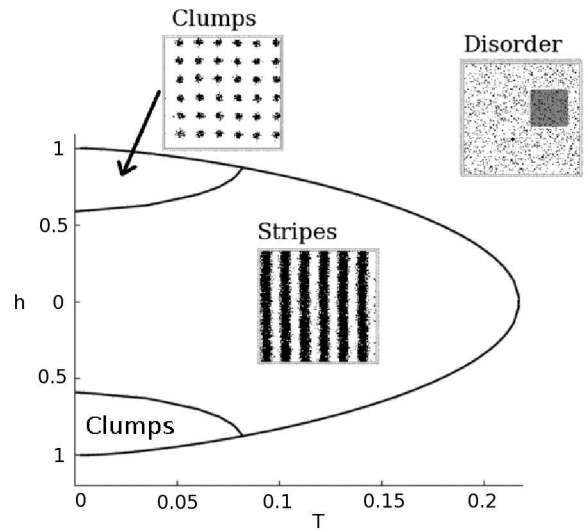}
\caption{Approximate phase diagram for the long-range antiferromagnetic Ising model with
a square shaped interaction. The system has two low temperature ordered phases, the clump phase and the stripe phase, and a disordered phase. The darkened square in the snapshot of
the disordered Ising model shows the interaction range for a spin
at the center of the square.\label{fig:squarePhaseDiagram}}
\end{figure}

We
study the two-dimensional long-range antiferromagnetic Ising
model with a square shaped interaction. For
this interaction a spin at $\boldsymbol{x}=(x,y)$ interacts with
a second spin at $\boldsymbol{x}'=(x',y')$ only if $|x-x'|\leq R$
and $|y-y'|\leq R$ (see \fig~\ref{fig:squarePhaseDiagram}). 

There are two types of periodic low temperature configurations, 
which we call the stripe phase and the clump phase, that are
energetically favorable with respect to the disordered phase. The system can lower its overall
interaction energy in the clump or stripe phases by grouping
like spins together. Although a spin pays an energy cost
due to its proximity to the other spins in its own clump or stripe,
it avoids interacting with most of the spins in neighboring clumps
or stripes. This system is related to a previously studied model
of particles~\citep{Klein1994}.

An approximate phase diagram is shown in \fig~\ref{fig:squarePhaseDiagram}.
The pictures are snapshots of the configurations from our Monte Carlo
simulations. From \fig~\ref{fig:squarePhaseDiagram} we see that the disorder-to-clumps transition
as well as the disorder-to-stripes transition are symmetry
breaking disorder-to-order transitions. In both cases the uniform
spatial symmetry is lost. The two order-to-order transitions are also
symmetry breaking. In the transition from stripes to clumps the continuous
translational symmetry along the direction of the stripes is lost.
In the clumps to stripes transition the $90^{\circ}$ rotational symmetry
is lost. 

The orientation of the square interaction restricts the possible alignments
of the stripes and clumps: stripes must be either horizontal
or vertical and clumps must be aligned in the vertical and horizontal
directions. A
circular-shaped potential which we do not consider here gives a honeycomb arrangement of
clumps and stripes with no angular restrictions. 

The Hamiltonian for the long-range antiferromagnetic Ising model is
\begin{equation}
H=-\frac{J}{z}\sum_{<ij>}\sigma_{i}\sigma_{j}-h\sum_{i}\sigma_{i},\label{eq:Hamiltonian}
\end{equation}
where the brackets indicate that the summation is over all spins within
the interaction range $R$, and $\sigma_{i}=\pm1$ refers to the value of a spin
at site $i$. The interaction $J=\pm1$,
where the positive sign corresponds to the ferromagnet and the negative
sign corresponds to the antiferromagnet; \begin{equation}
z=(2R+1)^{d}-1\label{eq:z}
\end{equation}
is the number
of spins within the interaction region of a single spin. Scaling the
strength of the interaction by the number of the spins in an interaction
range keeps the interaction energy per spin finite as
we increase the $R$~\citep{Kac1963}.

\subsection{Dissipative Langevin dynamics}

We now develop the equations for the relaxational Langevin dynamics as well as the background and fluctuation dynamics for a free energy which corresponds to the long-range antiferromagnetic Ising model. We begin with the coarse-grained free energy
\begin{align}
F[\phi(\boldsymbol{r})] & =  R^d \Big(-\frac{J}{2}\!\int d\boldsymbol{r}\left((\Lambda*\phi)(\boldsymbol{r})\right)\phi(\boldsymbol{r})\nonumber \\ & \quad -\!\int d\boldsymbol{r}\left(Ts[\phi(\boldsymbol{r})]+h\phi(\boldsymbol{r})\right)\Big),\label{eq:FreeEnergy}
\end{align}
where $\phi(\boldsymbol{r})=\dfrac{1}{(\Delta)^{d}}\sum_{i}S_{i}$ is the
coarse-grained magnetization of the spins inside a box of length $\Delta$
in $d$ dimensions centered at position $\boldsymbol{r}$. We use the notation $(\Lambda*\phi)(\boldsymbol{r})\equiv\!\int d\boldsymbol{r}'\Lambda(\boldsymbol{r}-\boldsymbol{r}')\phi(\boldsymbol{r}')$
to denote a spatial convolution. As in \eq~\eqref{eq:Hamiltonian} $J=+1$ for the ferromagnet and $J=-1$ for the antiferromagnet. The entropy density $s[\phi(\boldsymbol{r})]$ due to the coarse-graining
procedure is 
\begin{align}
s  [\phi(\boldsymbol{r})] & =  2\ln2-  \frac{1}{2}\{(1+\phi(\boldsymbol{r}))\ln{(1+\phi(\boldsymbol{r}))}\nonumber \\ & \qquad  +(1-\phi(\boldsymbol{r}))\ln{(1-\phi(\boldsymbol{r}))}\}.
\end{align}

The scaling of the strength of the interaction is implicit in the
interaction function $\Lambda(\boldsymbol{r})$ which we take to be a
repulsive pairwise step potential of the form \begin{equation}
\Lambda(x,y)=
\begin{cases}
1/z & (-R\leq x,y\leq R)\\
0 & \textnormal{otherwise}
\end{cases},
\end{equation}
where $z\sim (2R)^d$ for large $R$ (from \eq~\eqref{eq:z}).
This Kac form of the potential~\citep{Kac1963} yields a Hamiltonian that is well defined in the mean field limit; the energy
per spin remains finite as $R\rightarrow\infty$ because $\!\int d\boldsymbol{r}\Lambda(\boldsymbol{r})$
is independent of $R$. We believe that this form of the free energy
exactly represents the Ising model because a similar expression gives the exact
free energy for fluid density derived by Grewe and Klein~\citep{Grewe1977,Grewe1977a}. 

We can recover the Landau-Ginzburg free energy functional
in \eq~\eqref{eq:landau-ginsberg} by expanding $\Lambda(\boldsymbol{r}-\boldsymbol{r'})$ about the $q=0$ mode and neglecting terms higher than $O(q^2),$ and also expanding the 
entropy term in powers of $\phi$ and truncating terms higher than $O(\phi^4)$. For
the antiferromagnet the Fourier modes of interest are not centered about
$q=0$, so we will not truncate the potential.

For \dis\ Langevin dynamics the drift term $\Theta[\phi,\boldsymbol{r}]$ is proportional to
$-\delta F/\delta\phi$ times a mobility $M$,
which we are free to choose to correspond to the particular system
of interest. We allow the mobility to be spatially varying.
We take
\begin{equation}
\Theta[\phi,\boldsymbol{r}]=-M(\boldsymbol{r})\frac{\delta \tilde{F}}{\delta\phi(\boldsymbol{r})},\label{eq:dissipative_condition}
\end{equation}
where $\tilde{F}\equiv F[\phi(\boldsymbol{r})]/R^d$ is the free energy per interaction volume for large $R$ and is independent of $R$ since $F[\phi(\boldsymbol{r})]\propto R^d$ (\eq~\eqref{eq:FreeEnergy}). 
The equation of motion, \eq~\eqref{eq:generalLangevin}, becomes:
\begin{align}
\frac{\partial\phi(\boldsymbol{r},t)}{\partial t} & =  M(\boldsymbol{r})\Big\{J(\Lambda*\phi)(\boldsymbol{r},t) -\beta^{-1}\:\textnormal{arctanh}[\phi(\boldsymbol{r},t)] \nonumber \\
& \quad +h\Big\} +\sqrt{B(\boldsymbol{r})}R^{-d/2}\eta(\boldsymbol{r},t), \nonumber\\ &  \quad \mbox{(\dis\ Langevin dynamics)} \label{eq:relax_lang_dynamics}
\end{align}
where $B(\boldsymbol{r})=2M(\boldsymbol{r})\beta^{-1}$,
as determined by a fluctuation dissipation theorem.  We numerically solve \eq~\eqref{eq:relax_lang_dynamics} with
simulated noise in order to compare the relaxational Langevin dynamics with our predictions for stage two kinetics.

The dynamics in \eq~\eqref{eq:relax_lang_dynamics} evolves the system toward a local free energy minimum which is a solution to
the \el\ equation $\delta F/\delta\phi=0$
\begin{equation}
\phi_{0}(\boldsymbol{r})=\tanh\left[\beta J(\Lambda*\phi_{0})(\boldsymbol{r})+\beta h\right].\label{eq:ELIsing}\end{equation}

To compare to the Glauber dynamics in \sect~\ref{sec:glauber}, we now expand about the background configuration to find equations
for the evolution of the background configuration and the noise driven
fluctuations. Using the form of $\Theta[\phi,\boldsymbol{r}]$ given in \eq~\eqref{eq:dissipative_condition},
the nonlinear background dynamics in \eq~\eqref{eq:noiseless}
becomes
\begin{eqnarray}
\frac{\partial\phi_{b}(\boldsymbol{r},t)}{\partial t} & = & M(\boldsymbol{r})\Big(J(\Lambda*\phi_{b})(\boldsymbol{r},t)\nonumber \\
&&{} -\beta_{f}^{-1}\:\textnormal{arctanh}[\phi_{b}(\boldsymbol{r},t)]+h\Big),\label{eq:background_glauber}
\end{eqnarray}
and the linear theory for the fluctuations represented by \eq~\eqref{eq:generalLinear}
becomes
\begin{eqnarray}
\frac{\partial\psi(\boldsymbol{r},t)}{\partial t} & = & M(\boldsymbol{r})\left(J(\Lambda*\psi)(\boldsymbol{r},t)-\beta_{f}^{-1}\frac{\psi(\boldsymbol{r},t)}{1-\phi_{b}^2(\boldsymbol{r},t)}\right)\nonumber \\
&&{} + \sqrt{B(\boldsymbol{r})}\eta(\boldsymbol{r},t).\label{eq:dissipative_linear_theory}
\end{eqnarray}

\subsection{Glauber Langevin dynamics \label{sec:glauber}}

There is a Langevin equation corresponding to the long range Glauber
Monte Carlo dynamics which is not equivalent to the \dis\ Langevin dynamics of \eq~\eqref{eq:relax_lang_dynamics}. This Glauber Langevin equation, derived in Ref.~\citep{Barros},
is
\begin{eqnarray}
\frac{\partial\phi(\boldsymbol{r},t)}{\partial t} & = & \tanh\Big(\beta J(\Lambda*\phi)(\boldsymbol{r,}t)+\beta h\Big)-\phi(\boldsymbol{r},t) \nonumber \\
&&{} -\sqrt{B[\phi]}R^{-d/2}\eta(\boldsymbol{r},t),\label{eq:glauberlangevin}
\end{eqnarray}
where $B[\phi]=2-\tanh^{2}(\beta J(\Lambda*\phi)(\boldsymbol{r},t)+\beta h)-\phi^{2}(\boldsymbol{r},t)$
depends explicitly on $\phi(\boldsymbol{r},t)$ and the noise obeys
the conditions in Eqs.~\eqref{eq:scaled_noise_average} and~\eqref{eq:scaled_noise_correlation}.
 See also Ref.~\citep{Masi1996}.

Glauber dynamics obeys detailed balance and therefore should
take the system to equilibrium. We see that if we set the noise part
of the dynamics equal to zero the system converges to a solution of the \el\ equation
\begin{equation}
\frac{\partial\phi_{0}(\boldsymbol{r})}{\partial t}=0=\tanh(\beta J(\Lambda*\phi_{0})(\boldsymbol{r})+\beta h)-\phi_{0}(\boldsymbol{r}),\label{eq:ELglauber}
\end{equation}
in agreement with \eq~\eqref{eq:ELIsing}. 

Following a quench to $\beta_f$ and $h_f$ the linearized equation for Glauber dynamics (\eq~\eqref{eq:glauberlangevin}) from \eq~\eqref{eq:generalLinear}
is\begin{eqnarray}
\frac{\partial\psi(\boldsymbol{r},t)}{\partial t} & = & \Big(1-\tanh^{2}(\beta_{f}J(\Lambda*\phi_{b})(\boldsymbol{r})+\beta_{f}h)\Big) \nonumber \\
&&{} \times J\beta_{f}(\Lambda*\psi)(\boldsymbol{r},t)-\psi(\boldsymbol{r},t)\label{eq:glauber_linear} \nonumber \\
&&{} -\sqrt{B\vert_{\phi=\phi_{b}}}\eta(\boldsymbol{r},t).\end{eqnarray}
When $\phi_b(\boldsymbol{r},t)\simeq\phi_{0}(\boldsymbol{r})$ we can apply \eq~\eqref{eq:ELglauber} to \eq~\eqref{eq:glauber_linear}
and obtain an expression for stage two kinetics\begin{eqnarray}
\frac{\partial\psi(\boldsymbol{r},t)}{\partial t} & = & \left(1-\phi_{0}^{2}(\boldsymbol{r})\right)\beta_{f}J(\Lambda*\psi)(\boldsymbol{r},t)-\psi(\boldsymbol{r},t) \nonumber \\
&&{} -\sqrt{2(1-\phi_{0}^{2}(\boldsymbol{r}))}\eta(\boldsymbol{r},t).\label{eq:chc-like_glauber}
\end{eqnarray}
Equation~\eqref{eq:chc-like_glauber}
for Glauber Langevin stage two dynamics is identical to \eq~\eqref{eq:dissipative_linear_theory} for the \dis\ Langevin
dynamics of stage two if the mobility is chosen to be \begin{equation}
M(\boldsymbol{r})=\beta_{f}(1-\phi_{0}^2(\boldsymbol{r}))\label{eq:mobility}.
\end{equation} 

\section{disorder-to-order transitions\label{sec:disorder-to-order}}

\subsection{Theory}

We first treat the transition
from the disordered phase to either stripes or clumps in the antiferromagnetic Ising model. We consider \dis\ dynamics (\eq~\eqref{eq:glauberlangevin}) where we choose the mobility to obey \eq~\eqref{eq:mobility}, where $\phi_0(\boldsymbol{r})=\phi_0$ so that $M(\boldsymbol{r})=M$. For this choice, the \dis\ dynamics and the Glauber dynamics should be equivalent for stage two.  

 For comparison we will also discuss the contrasting behavior of the ferromagnetic Ising model. In this case
the initial background configuration is uniform and $\phi_{b}(t=0)$ may or may not be a solution to the \el\
equation. Consider the uniform solutions to the
\el\ equation (\eq~\eqref{eq:ELIsing}) given by \begin{equation}
\phi_{0}=\tanh\left(\beta(J\phi_{0}+h)\right),\label{eq:soln}
\end{equation}
where we have used $\int d\boldsymbol{r}\Lambda(\boldsymbol{r})=1$.
For $h=0$, $\phi_0=0$ is a stable minimum for the antiferromagnet and the
ferromagnet at high temperatures. For the ferromagnet at low temperatures
and relatively small $|h|$, there is one solution unstable to $q=0$ fluctuations and two
solutions, either both stable, or one stable and one metastable.

For the antiferromagnet at low temperatures and relatively small $|h|$
we know that the stable solution is not spatially uniform. The uniform unstable solution given by \eq~\eqref{eq:soln} is stable for the $q=0$ mode and unstable for certain $q>0$ modes. 
The details of the instability are discussed in the following. 

The linearized equation for stage one kinetics is
of the form of \eq~\eqref{eq:dissipative_linear_theory}, with $\phi_b(\boldsymbol{r},t)=\phi_b(t)$. Equation~\eqref{eq:dissipative_linear_theory} can be written in Fourier space as  
\begin{equation}
\frac{\partial\psif(\boldsymbol{q},t)}{\partial t}=D(\boldsymbol{q},t)\psif(\boldsymbol{q},t)+\sqrt{B}\eta(\boldsymbol{q},t),\label{eq:Fourier_stage1}
\end{equation}
where
\begin{equation}
D(\boldsymbol{q},t)  =  M\left(J\Lambda(\boldsymbol{q})-\frac{1}{\beta_{f}(1-\phi_{b}^{2}(t))}\right).
\end{equation}
The solution to \eq~\eqref{eq:Fourier_stage1} is not an exponential in general, but can be solved numerically by simultaneously solving
\eq~\eqref{eq:background_glauber} for $\phi_{b}(t)$.

Eventually, we expect that $\phi_{b}(t)\simeq\phi_{0}$ and that stage
two kinetics takes over. The dynamics of the fluctuations is then given by Eq.~\eqref{eq:Fourier_stage1} with $\phi_b(\boldsymbol{r},t)=\phi_0$, so that $D(\boldsymbol{q},t)$ for stage two kinetics becomes\begin{equation}
D(\boldsymbol{q})=\beta_f(1-\phi_{0}^{2})J\Lambda(\boldsymbol{q})-1.\label{eq:DII}
\end{equation}
With these simplifications, we solve Eq.~\eqref{eq:Fourier_stage1} to obtain a generalization of \eq~\eqref{eq:struct_factor} for disorder-to-order transitions
\begin{eqnarray}
S(\boldsymbol{q},t) & = & \frac{\delta(\boldsymbol{q)}^2\phi_0^2}{V}+ \Big[e^{2D(\boldsymbol{q})t}\left(S_0(\boldsymbol{q})+\frac{1-\phi_{0}^{2}}{D(\boldsymbol{q})}\right) \nonumber \\ &&{}  -\frac{1-\phi_{0}^{2}}{D(\boldsymbol{q})}\Big],\label{eq:strcfunc}
\end{eqnarray}
where $S_0(\boldsymbol{q})=\left\langle \vert \psi(t=t_0)\vert^2\right\rangle /V$ is determined by the initial conditions for the differential equation~\eqref{eq:Fourier_stage1} and can be treated as a fitting parameter unless stage two kinetics begins immediately after the quench where $S_0(\boldsymbol{q})=S(\boldsymbol{q},t=0)$. Equation~\eqref{eq:strcfunc} is only valid for stage two kinetics, that is, for $t>t_0$. 

\begin{figure}[h]
\subfigure[\label{fig:Dk.a}]{
\includegraphics[scale=0.75]{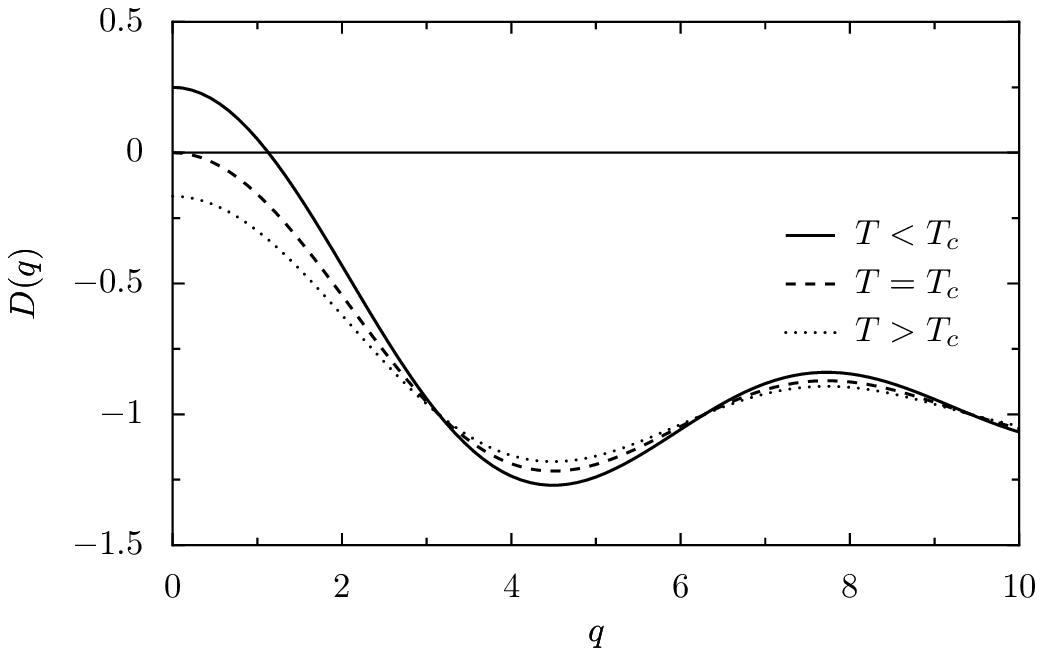}}
\quad
\subfigure[\label{fig:Dk.b}]{
\includegraphics[scale=0.75]{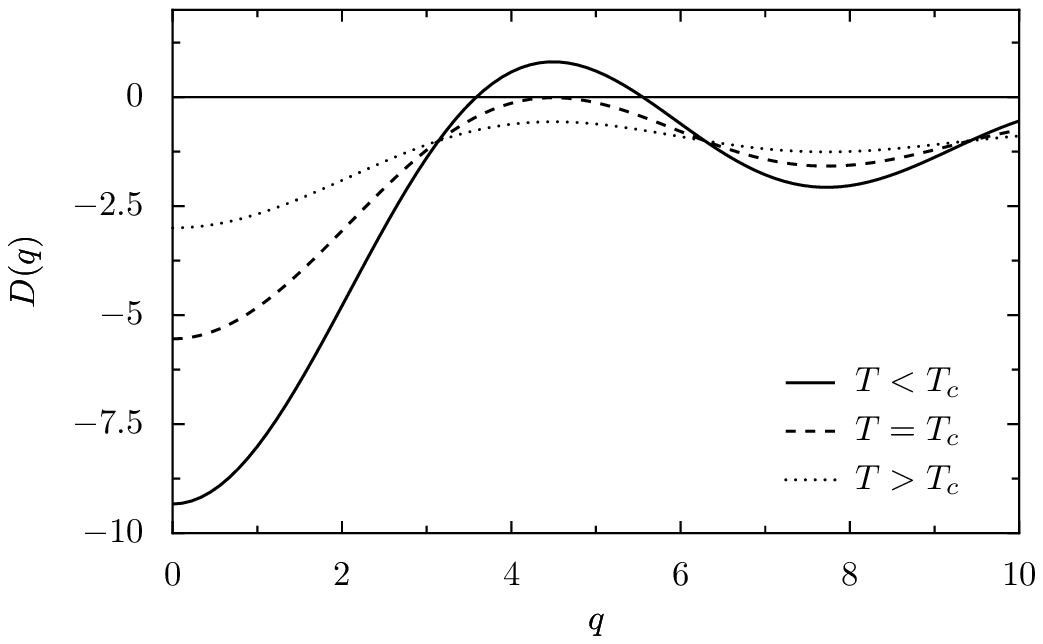}}
\caption{The function $D(q_x,0)$ is plotted for (a) the ferromagnet
($J=1$) and the (b) antiferromagnet ($J=-1$) for various temperatures. Using Eqs.~\eqref{eq:DII} and~\eqref{eq:potential} the
Fourier modes corresponding to positive values of $D(q)$ are
exponentially growing. As the temperature is lowered, the ferromagnet
first becomes unstable at $q=0$, but the antiferromagnetic becomes
unstable at $q=q^{*}\simeq4.4934$ (see \eq~\eqref{eq:DII}). \label{fig:Interaction-potential-in}}
\end{figure}

The sign of $D(\boldsymbol{q})$ in Eq.~\eqref{eq:DII} determines whether the structure
function grows or decays exponentially. To see for which
$\boldsymbol{q}$ value $D(\boldsymbol{q})$ first changes from negative to positive as the temperature is lowered to less than $T_{c}$,
we look at the potential $\Lambda$ in Fourier space. For the square
shaped interaction we have
\begin{equation}
\Lambda(q_{x},q_{y})=\frac{\sin(q_{x})}{q_{x}}\frac{\sin(q_{y})}{q_{y}}. \label{eq:potential}\end{equation}
We plot the function $D(q_x,0)$ in \fig~\ref{fig:Interaction-potential-in}. 

For the ferromagnet ($J=+1$) $D(\boldsymbol{q})$ for $\boldsymbol{q}=0$ first
crosses from negative to positive as the temperature is lowered. As the temperature is lowered
for the antiferromagnetic case ($J=-1$), $D(q)$
first crosses from negative to positive at the global minimum of
$\Lambda(\boldsymbol{q})$ which occurs at $\boldsymbol{q}=(\pm q^*,0)$ or $\boldsymbol{q}=(0,\pm q^*)$ where $q^{*}\simeq4.4934$. The temperature
at which this crossover occurs is the critical temperature $T_{c}$ for a given $\phi_0$. 
The critical temperature for the antiferromagnetic system is
\begin{equation}
T_{c}=-\frac{\sin(q^{*})}{q^{*}}(1-\phi_{0}^{2}).\label{eq:critTemp}
\end{equation}
Below this
temperature there is at least one unstable mode which grows exponentially.

We find from numerical solutions of \eq~\eqref{eq:relax_lang_dynamics} that the disorder-to-order and order-to-disorder transitions occur at the same temperature, indicating the existence of a continuous phase transition. Thus the line in the phase diagram (Fig.~\ref{fig:squarePhaseDiagram}) separating the disordered phase from the ordered phases is a line of critical points.

\subsection{Simulation Results}

\subsubsection{Critical Quench}

\begin{figure}
\includegraphics[scale=0.75]{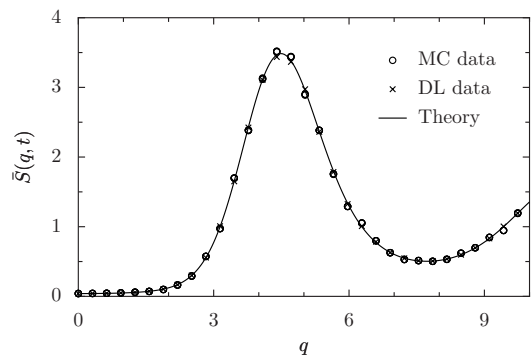}
\caption{Structure factor $\bar{S}(q,t)\equiv [S(\pm q,0;t)+S(0,\pm q;t)]/4$ versus $q$ for the antiferromagnetic Ising model for early time $t=0.125$ MCS after a critical quench from $T_i=\infty$ tp $T_f=4T_c/9.$
The Monte Carlo data (MC data) has an interaction
range of $R=128$, while the numerical solutions to the dissipative Langevin \eq~\eqref{eq:relax_lang_dynamics} (DL data) has
a range of $10^{6}$. The theory line is a plot of \eq~\eqref{eq:strcfunc}.\label{cap:Structure-factor-versus}}
\end{figure}

We first consider the critical quench for the antiferromagnet for which we expect stage two
kinetics immediately following the quench (refer to Table~\ref{tab:summary_table}). For this case 
$M=\beta_{f}$ from \eq~\eqref{eq:mobility}.
Figure~\ref{cap:Structure-factor-versus}
shows the correspondence between the linear theory prediction \eq~\eqref{eq:strcfunc}, the
numerical solutions to the \dis\ Langevin equation, \eq~\eqref{eq:relax_lang_dynamics},
averaged over the noise, and the Monte Carlo simulations averaged
over the noise. In \fig~\ref{cap:Structure-factor-versus} the structure factor is plotted versus
$q$ at an early time $t=0.125$ MCS (Langevin time is measured in units of MCS). Because we use a square interaction,
we plot
\begin{equation}
\bar{S}(q,t)\equiv [S(\pm q,0;t)+S(0,\pm q;t)]/4\end{equation}
instead of taking a circular average. For all simulations
presented, we hold the ratio of the system size to the interaction
range constant at $L/R\simeq2.78$ so that the stripe state
of the final phase is approximately commensurate with the system size~\cite{finite}.

The coarse-graining employed in the Langevin equation allows us to consider
system sizes and interaction ranges which are much larger than those of the Monte Carlo
simulations. In \fig~\ref{cap:Structure-factor-versus} the Monte Carlo simulations have $R=128$, and
the Langevin solutions have $R=10^{6}$.
The theory curve corresponds to an infinite interaction range (so that the linear
theory lasts forever.) Figure~\ref{cap:Structure-factor-versus} shows that the linear theory is still a
good approximation at time $t=0.125$ for the Langevin and
the Monte Carlo dynamics. 

\begin{figure}
\includegraphics[scale=0.75]{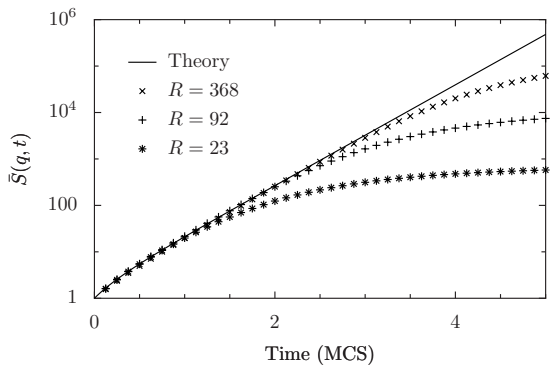}
\caption{Semi-log plot of the structure factor $\bar{S}(q,t)$ for $q=4.516\simeq q^*$ versus time for our Monte Carlo
simulations of a critical quench in the antiferromagnetic Ising model. The temperature is quenched from
$T_{i}=\infty$ to $T_{f}= 4T_{c}/9$. The theory line is a plot of \eq~\eqref{eq:strcfunc}. The duration
of the exponential growth increases as the interaction range increases.
\label{fig:range compare}}
\end{figure}

Next we compare the time dependence of $S(\boldsymbol{q},t)$ given by \eq~\eqref{eq:strcfunc}
following a critical quench to our Monte Carlo simulations. Figure~\ref{fig:range compare}
shows $\bar{S}(q,t)$ for $q\simeq q^{*}$ plotted
against time for several values of $R$ along with the
theoretical prediction of \eq~\eqref{eq:strcfunc}.  As predicted by Binder~\citep{Binder1984}, we find that
stage two kinetics holds for longer times as we increase the interaction
range for this value of $q$. 

\subsubsection{Off-critical quenches}

\begin{figure}
\includegraphics[scale=0.75]{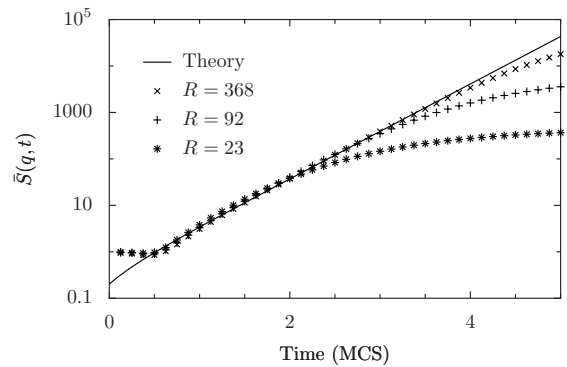}
\caption{Semi-log plot of the structure factor $\bar{S}(q,t)$ for $q=4.516\simeq q^*$ versus time for Monte Carlo
simulations of an off-critical quench in the antiferromagnetic Ising model. The simulation parameters are $T_{i}=\infty$,
$T_{f}=\frac{4}{9}T_{c}$, $h_{i}=0$, and $h_{f}=0.5$. The first Monte Carlo step is stage one kinetics, whose time scale is independent of interaction
range. The duration of stage two kinetics, marked by exponential evolution,
increases as the interaction range increases. The theory line is a
plot of \eq~\eqref{eq:strcfunc} (with $S_0(q)$ fit to data) and
is only expected to be applicable for stage two. \label{cap:off-crit}}

\end{figure}

For external field quenches or off-critical temperature quenches we find that
the short time behavior is independent of $R$, and
corresponds to our prediction of stage one kinetics. In \fig~\ref{cap:off-crit}
we present the results of off-critical quenches ($T_{i}=\infty$,
$h_{i}=0$, $T_{f}=4T_{c}/9$, $h_{f}=0.5$) for $R=23,\:92,\:368$ for Monte Carlo simulations
averaged over at least 700 independent initial configurations for each value of $R$. We plot $\bar{S}(q,t)$ versus time for $q=4.516\simeq q^*$. We interpret the first Monte Carlo step to be stage
one kinetics, which, as we see from \fig~\ref{cap:off-crit}
has a time scale that is independent of the interaction range. At
about one Monte Carlo step stage two kinetics take over, and the structure factor
at $q\simeq q^{*}$ grows exponentially until stage two kinetics
breaks down. The break down occurs later for longer interaction ranges.
These observations are consistent with the prediction that the time
scale of stage one kinetics is determined by the noiseless dynamics of \eq~\eqref{eq:noiseless} which is independent of $R$. Note that stage one kinetics were not predicted by the \chc\ theory.

\begin{figure}[h]
\subfigure[\label{fig:disorder_to_stripes}]{
\includegraphics[scale=0.75]{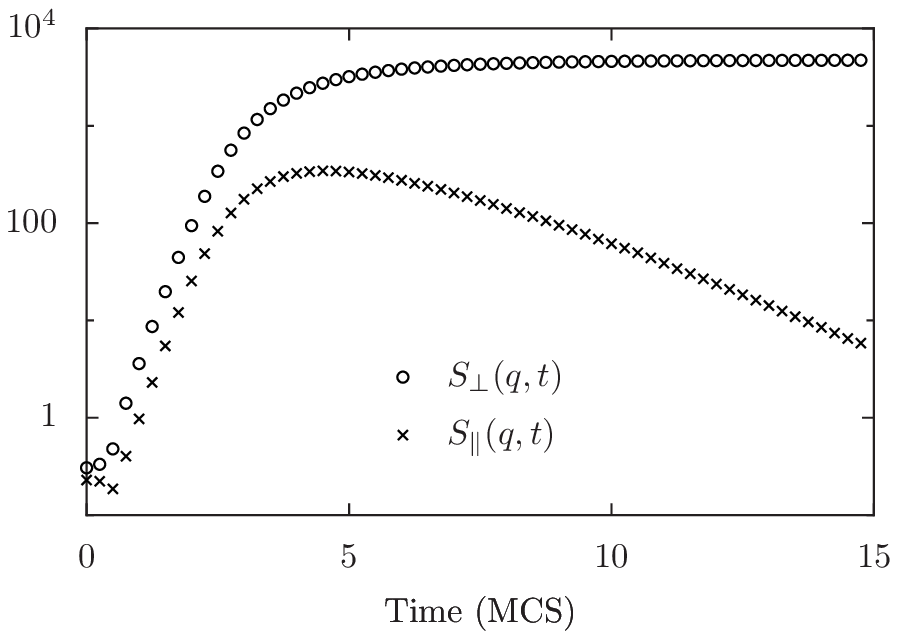}}
\quad
\subfigure[\label{fig:disorder_to_clumps}]{
\includegraphics[scale=0.75]{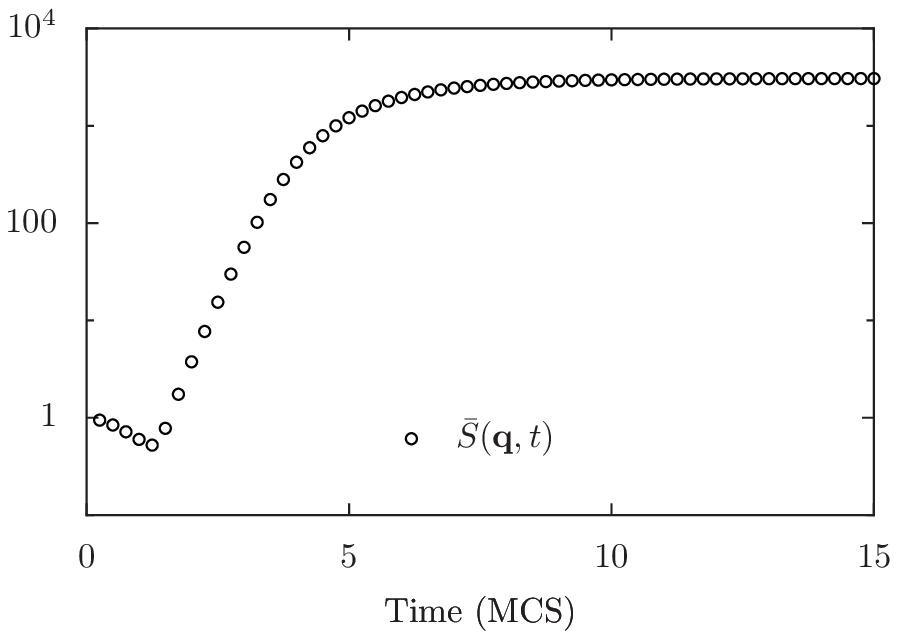}}
\caption{(a.)A semi-log plot of $S_{\perp}(q,t)$ (where $\boldsymbol{q}$ is in the direction perpendicular to the final stripe phase) and $S_{\parallel}(q,t)$ (where $\boldsymbol{q}$ is in the direction parallel to the final stripe phase) versus time for an off-critical quench from disorder to stripes with $T_f=0.07,$ $h_f=0.4$ and $R=92.$  Stage two lasts until about 1.5 MCS, after which $S_{\perp}(q,t)$ and $S_{\parallel}(q,t)$ begin to separate and eventually $S_{\parallel}(q,t)$ decreases towards its stable value. (b.) A semi-log plot of $\bar{S}(q,t)$ for an off-critical quench from disorder to clumps with $T_f=0.02,$ $h_f=0.8$ and $R=92$ is plotted for comparison. \label{fig:parallel_perp}}
\end{figure}

To show that our generalized theory predicts the same behavior for both the disorder to stripes and the disorder to clumps transitions, we compare $S_{\perp}(q,t)$ ($\boldsymbol{q}$ in the direction perpendicular to the final stripe phase) and $S_{\parallel}(q,t)$ ($\boldsymbol{q}$ in the direction parallel to the final stripe phase) where $S_{\perp}(q,t)=S(q^*,0;t)$ and $S_{\parallel}(q,t)=S(0,q^*;t)$ for vertical stripes and $S_{\perp}(q,t)=S(0,q^*;t)$ and $S_{\parallel}=S(q^*,0;t)$ for horizontal stripes.  We find that $S_{\perp}(q,t)$ and $S_{\parallel}(q,t)$ both grow exponentially with the same rate during stage two (see Fig~\ref{fig:disorder_to_stripes}).  It is not until after stage two that $S_{\parallel}(q,t)$ begins to decrease toward its final value. For comparison, $\bar{S}(q,t)$ is plotted in Fig.~\ref{fig:disorder_to_clumps} for the disorder to clumps case. Note that the qualitative behavior for the disorder to stripe and the disorder to clump transitions are the same.

\section{stripes to clumps transition\label{sec:order-to-order}}

We now consider the unstable kinetics for spatial symmetry
breaking order-to-order transitions. In this section we develop a
theory that describes the early time kinetics from an initial stripe
state to the stable clump phase following an instantaneous quench
of the external field. The development
of the theory is analogous to that of the disorder-to-order case, but the periodic structure of the initial configuration of the
system introduces some interesting deviations from the \chc\ theory. 

\subsection{Theory}

After an external field quench with the temperature held constant, the
stable solution to the \el\ equation is a clump configuration corresponding
to $T_f$ and $h_f$ (see Fig.~\ref{fig:squarePhaseDiagram}). We find from numerical solutions to \eq~\eqref{eq:relax_lang_dynamics} 
that there is a second, unstable stripe solution as well, $\phi_{s}(\boldsymbol{r})=\phi_{s}(r_x)$ where $r_x$ and $r_y$ represent scaled spatial coordinates
(we assume here that the stripes are in the vertical direction). 

As in \sect~\ref{sec:disorder-to-order}, we proceed by considering \dis\ dynamics (\eq~\eqref{eq:glauberlangevin}) where we choose the mobility to obey \eq~\eqref{eq:mobility}, where we now choose $\phi_0(\boldsymbol{r})=\phi_s(r_x)$ so that $M(\boldsymbol{r})=M(r_x)$. Again, the \dis\ dynamics and the Glauber dynamics should be equivalent for stage two.  

The
nonlinear dynamics of \eq~\eqref{eq:noiseless} evolves the initial
stripe configuration ($\phi_{b}(r_x,r_y;t=0)=\phi_{b}(r_x;t=0)$) toward
$\phi_{s}(r_x).$ When $\phi_{b}(r_x,t)\simeq\phi_{s}(r_x)$, stage two kinetics begins.
In this case stage two kinetics is given by \eq~\eqref{eq:dissipative_linear_theory} where the mobility $M(r_x)$ is given by \eq~\eqref{eq:mobility} with $\phi_0(\boldsymbol{r})=\phi_s(r_x)$ and $J=-1.$ We find that in order to determine the time dependence of the structure factor for stage two it is sufficient to consider the dynamical equation~\eqref{eq:dissipative_linear_theory} without the noise term:
\begin{equation}
\frac{\partial\psi(\boldsymbol{r},t)}{\partial t}=-M(r_x)(\Lambda*\psi)(\boldsymbol{r},t)-\psi(\boldsymbol{r},t).\label{eq:stripes-clumps-Langevin}
\end{equation}
To understand the behavior of the Fourier modes we Fourier transform \eq~\eqref{eq:stripes-clumps-Langevin}:
\begin{eqnarray}
\frac{\partial\psif(q_{x},q_{y};t)}{\partial t} & = & -\!\int dq_x' M(q_x'-q_{x})\Lambda(q_x',q_{y})\psif(q_x',q_{y};t)\nonumber \\
&& -\psif(q_{x},q_{y};t).\label{eq:fSpace}
\end{eqnarray}
Note that in \eq~\eqref{eq:fSpace} the $q_{y}$ modes (along the direction of the stripes) are
decoupled, but the $q_{x}$ modes (perpendicular to the direction
of the stripes) are not. Hence the Fourier modes do not
have purely exponential growth as in the disorder-to-order case.

To solve \eq~\eqref{eq:fSpace} numerically,
we discretize space. For a lattice with $L\times L$ coarse-grained bins, and therefore
$L\times L$ Fourier modes, we write $\psi(q_{x},q_{y})$ as $\psi_{ij}$, where $q_{x}$ and $q_{y}$ are $q_{x}=2\pi i/L$ and $q_{y}=2\pi j/L$
for $0\leq i,j<L$. Because the $q_{y}$ variables are decoupled,
we choose a particular value $j=j_{0}$ for the remainder of the calculation.
The discretized \eq~\eqref{eq:fSpace} can be written in matrix
form
\begin{equation}
\frac{\partial\psif_i(t)}{\partial t}=A_{ik}\psif_k(t),\label{eq:VecEq}
\end{equation}
where we have left off the $j=j_0$ index.
The matrix $A$ is real and has elements
\begin{equation}
A_{ik}=-M_{ik}\Lambda_{k}-\delta_{ik}.\label{eq:matrix}\end{equation}
To solve the homogeneous set of equations, $\partial\psif_i(t)/\partial t=A_{ik}\psif_k(t),$ we diagonalize $A$ to find eigenvectors $v^{\alpha}$
which satisfy:
\begin{equation}
A_{ik}v_k^{\alpha} =\lambda^{\alpha} v_i^{\alpha},\label{eq:eigen}
\end{equation}
where the index $\alpha = 1,\,2,\,...\,,\,L.$ If the eigenvalues $\lambda^{\alpha}$ are real and distinct (which we have found to be true for the matrices we have numerically diagonalized) then $v^{\alpha}$ form a new basis.
We expand $\psif$ in this basis
\begin{equation}
\psif_i(t)=\sum_{\alpha}a^{\alpha}(t)v_i^{\alpha}.\end{equation}
Equation~\eqref{eq:VecEq} becomes\begin{equation}
\frac{\partial}{\partial t}\sum_{\alpha}a^{\alpha}(t)v_i^{\alpha}=\sum_{\alpha}a^{\alpha}(t)\lambda^{\alpha}v_i^{\alpha}.\label{eq:decompEq}\end{equation}
Because $v^{\alpha}$ are linearly independent, \eq~\eqref{eq:decompEq} can be written as a set of equations:\begin{equation}                                                                                                        \frac{\partial}{\partial t}a^{\alpha}(t)=a^{\alpha}(t)\lambda^{\alpha},                                                                                                   \end{equation}
with solutions
\begin{equation}
a^{\alpha}(t)=a^{\alpha}(0)\textnormal{e}^{\lambda^{\alpha}t}.
\end{equation}
The structure factor $S(q_x=2\pi i/L, q_y=2\pi j_0/L;t)=S_{ij_0}(t)=\left\langle \vert \phi_{ij_0}(t)\vert^2\right\rangle /V$ due to the noiseless dynamics is \begin{align}
S_{ij_0}(t) & =  \frac{\phi_{s,i}^2(\delta_{0,i})^2}{V} \nonumber \\ & \quad +\sum_{\alpha,\alpha'}\frac{\left\langle a^{\alpha}(0)a^{\alpha'}(0)\right\rangle}{V}e^{(\lambda^{\alpha}  +\lambda^{\alpha'})t} v_{ij_0}^{\alpha}v_{ij_0}^{\alpha'}, \label{eq:OOStructureFactor}
\end{align}
where $\phi_{s,i}=\phi_s(q_x=2\pi i/L)$ and we expect that $\left\langle a^{\alpha}(0)a^{\alpha'}(0)\right\rangle\sim V.$
In the unstable regime, there is at least one positive $\lambda^{\alpha}$ indicating exponential growth. From \eq~\eqref{eq:OOStructureFactor} we see the exponential growth lasts throughout stage two, but it is growth of eigenvectors that will, in general, mix Fourier modes. We expect, however, that after some
time the term in the sum in \eq~\eqref{eq:OOStructureFactor} which corresponds to the
largest positive eigenvalue will dominate, and the growth of Fourier
modes will be well approximated by pure exponential growth. Note that we only expect to see exponential growth for near-mean-field systems whose long interaction ranges extend the lifetime
of the stage two kinetics to $\tau\sim\ln R$ at the fastest growing
eigenvector. 

\subsection{Simulation results}

\subsubsection{External field quench\label{sub:External-field-quench}}

\begin{figure}
\begin{tabular}{cc}
\includegraphics[width=3cm]{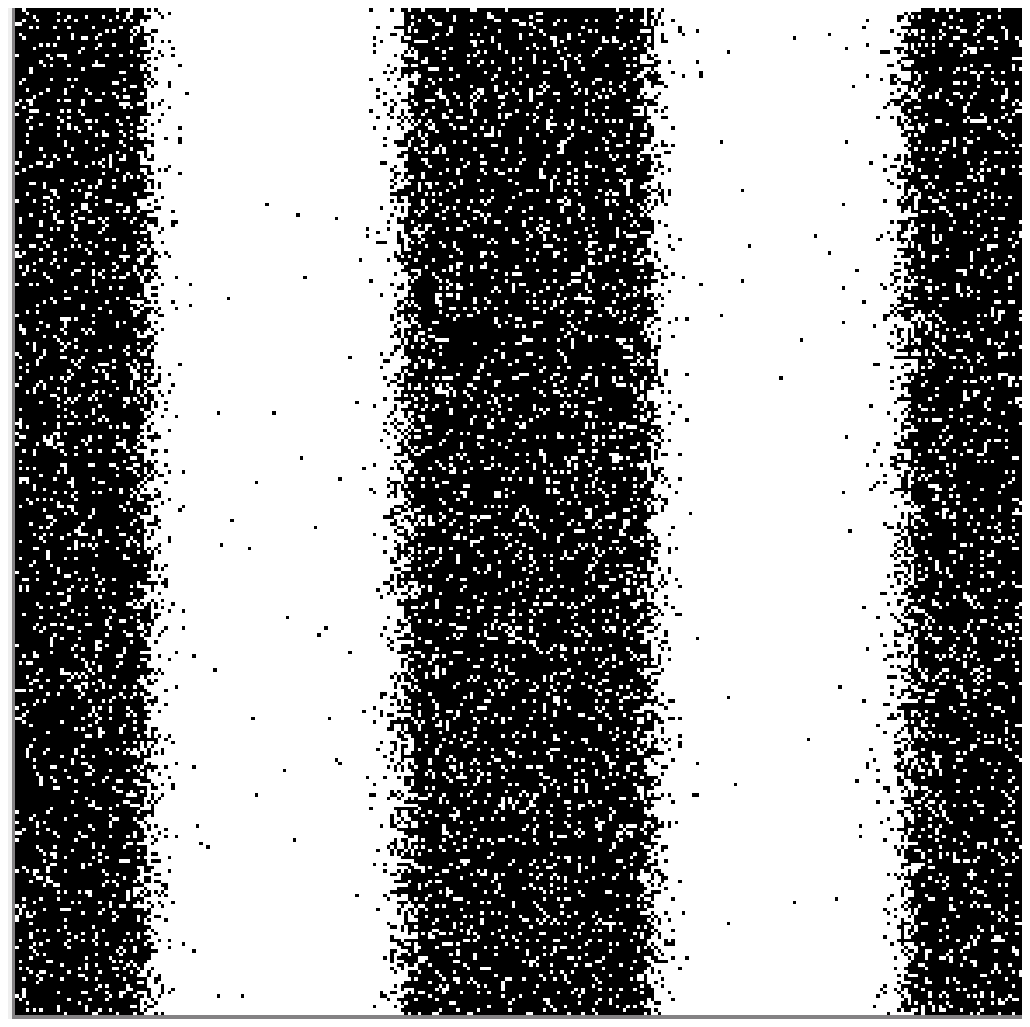} & \includegraphics[width=3cm]{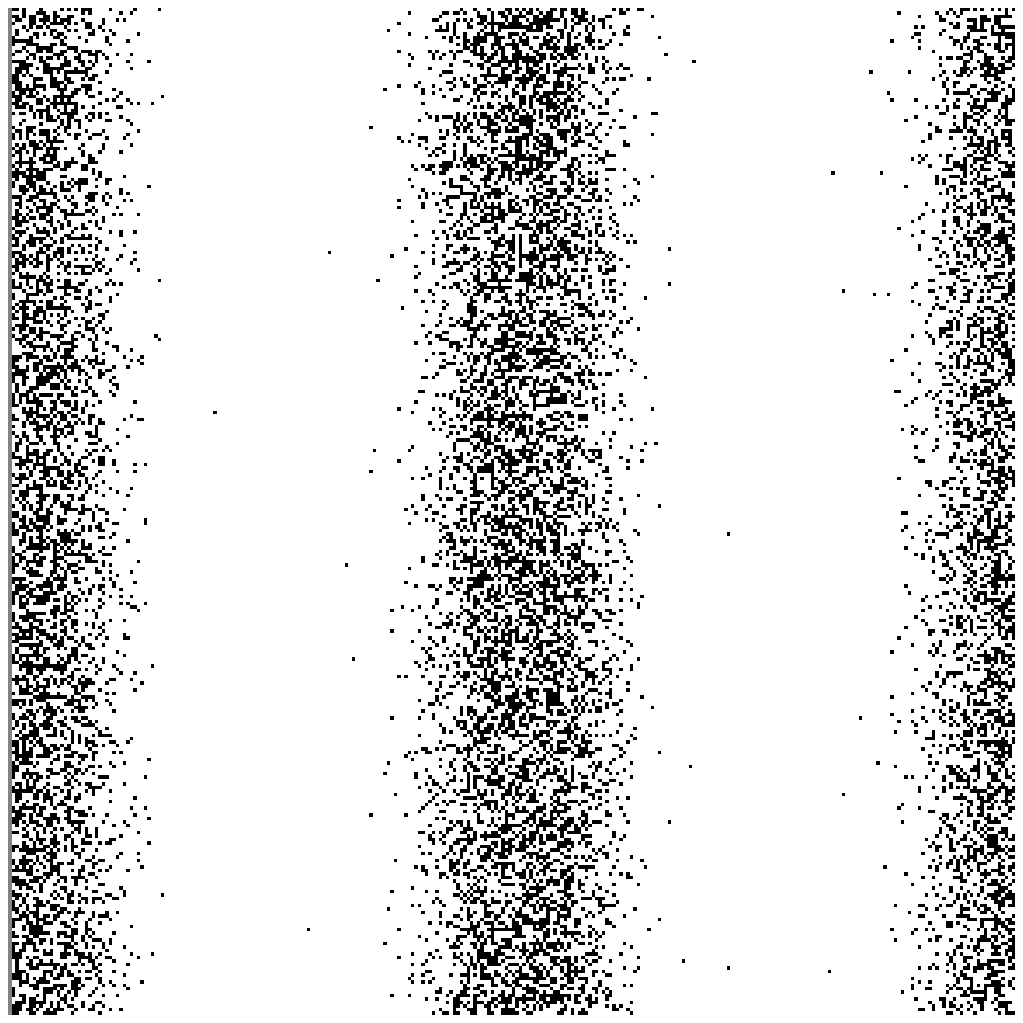}\tabularnewline
t=0.125 & t=5.00\tabularnewline
\includegraphics[width=3cm]{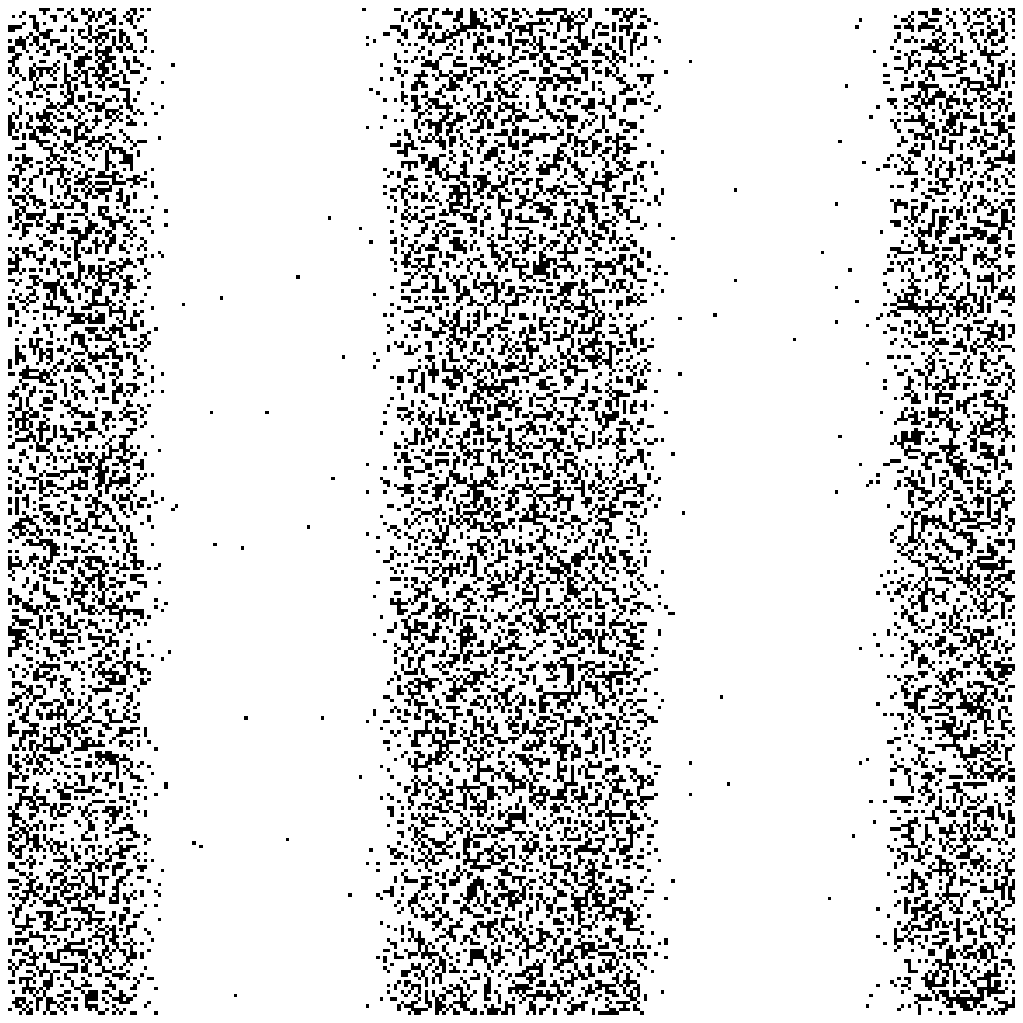} & \includegraphics[width=3cm]{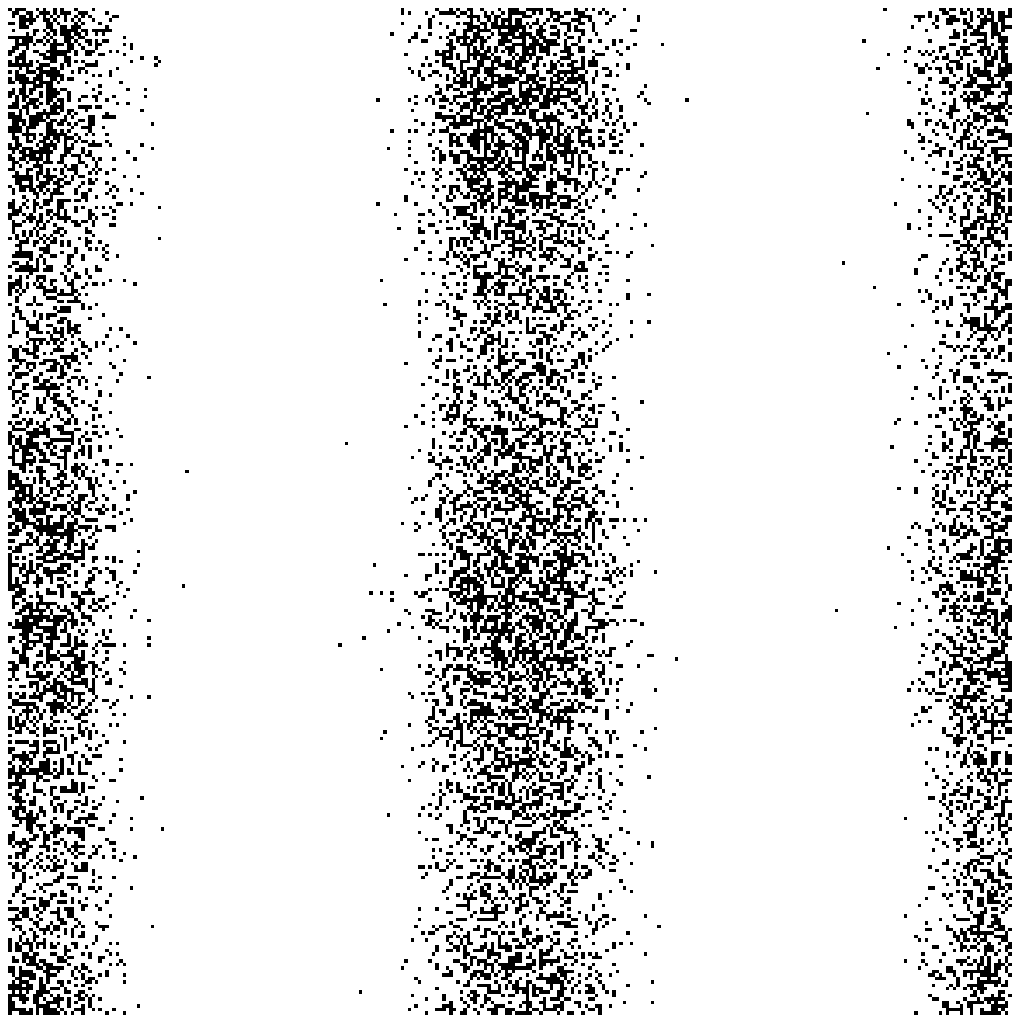}\tabularnewline
t=1.00 & t=10.0\tabularnewline
\includegraphics[width=3cm]{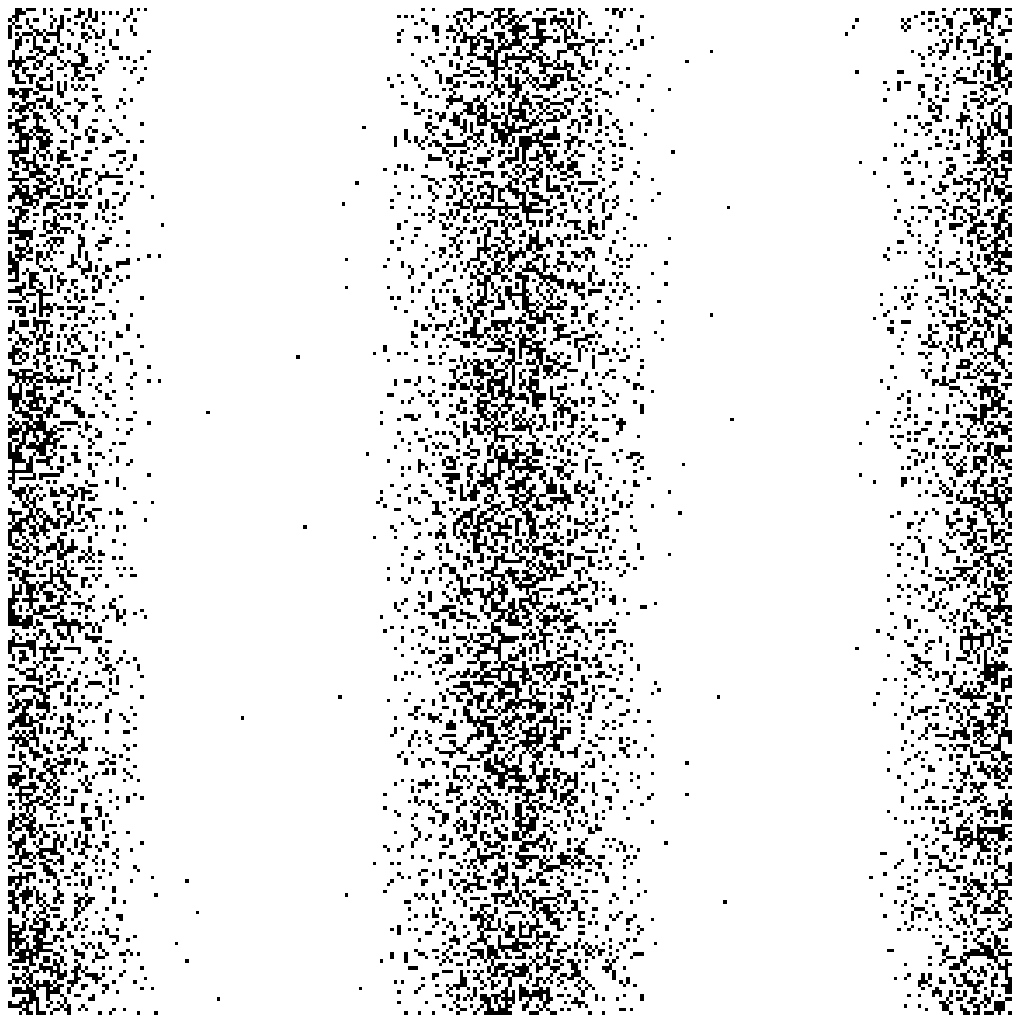} & \includegraphics[width=3cm]{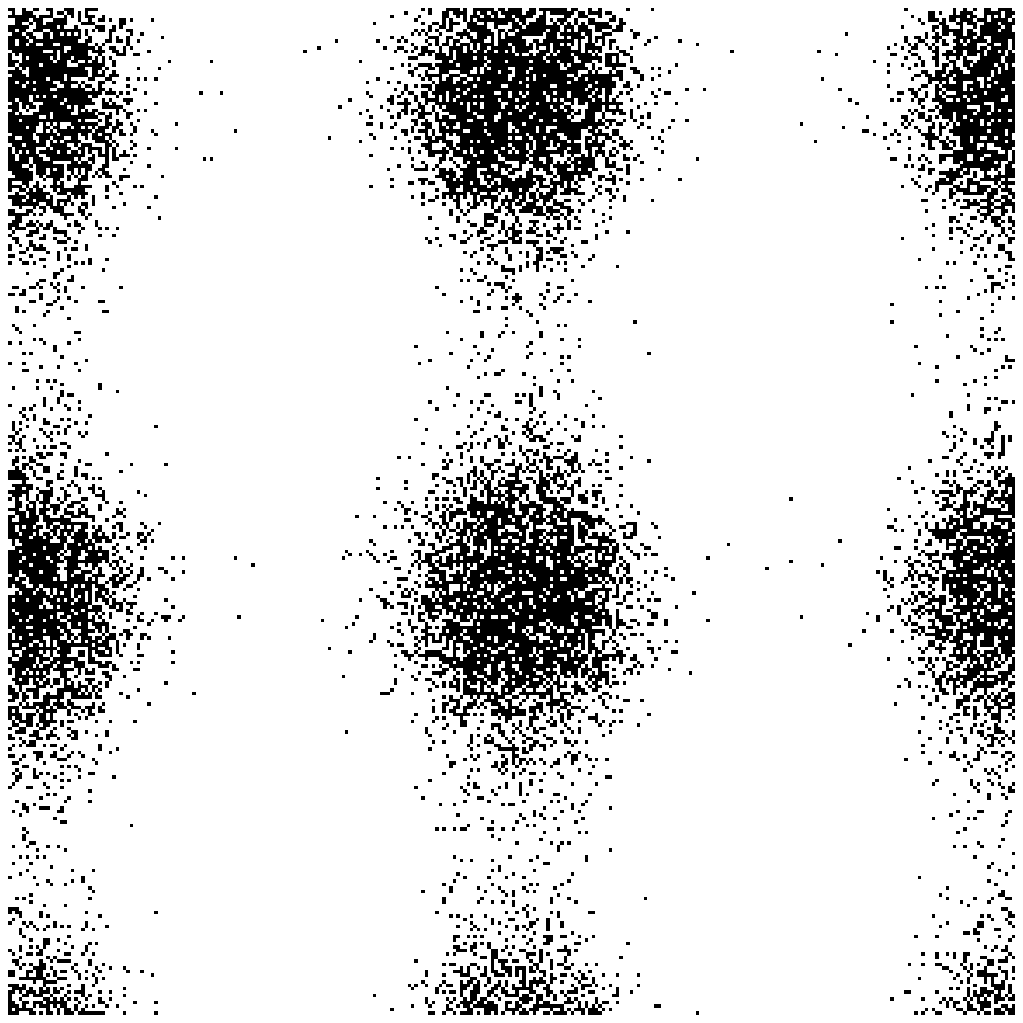}\tabularnewline
t=2.50 & t=15.0\tabularnewline
\end{tabular}
\caption{Several snapshots showing the initial evolution from an unstable stripe
configuration to a clump configuration in the antiferromagnetic Ising model. The lattice size is 1024,
the interaction range is 368, and $T_{i}=T_{f}=0.04$. Time is measured
in Monte Carlo steps beginning after the external field quench from
$h_{i}=0$ to $h_{f}=0.8$. The corresponding structure factor is shown in \fig~\ref{fig:Plot-of-structure}.\label{fig:OO-snapshots}}
\end{figure}

We present the results of Monte Carlo simulations for a quench from
a stable stripe configuration with $h_{i}=0$ to $h_{f}=0.8$ at $T_{i}=T_{f}=0.04$.
Snapshots following this quench are shown in \fig~\ref{fig:OO-snapshots}
for a system with $L=1024$ and interaction range $R= 368$.
We find that the stripes quickly become less dense and then narrower.
Finally, around 10\,MCS, the stripes begin to form clumps, and
the modulation along the length of the clumps becomes visually perceptible. 

\begin{figure}
\includegraphics[scale=0.75]{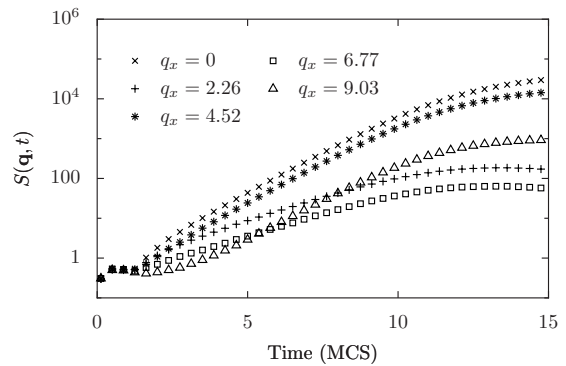}
\caption{Semi-log plot of the structure factor $S(\boldsymbol{q},t)$ versus time in the antiferromagnetic Ising model with $R=368$ for various values of $q_x$
for $q_{y}\simeq q^{*}$ following a quench from $h_{i}=0$
to $h_{f}=0.8$ at $T_{i}=T_{f}=0.04$.
The Monte Carlo data is averaged over about 700 independent
stripe configurations. We interpret the first Monte Carlo step
to be stage one kinetics. After this time stage two takes over. At the
beginning of stage two, the Fourier modes do not grow exponentially.
Eventually they grow exponentially at two different
rates (starting from about 4 Monte Carlo steps). This exponential growth lasts until about 10 Monte Carlo steps for $R=368$.
\label{fig:Plot-of-structure}}
\end{figure}

In Fig.~\ref{fig:Plot-of-structure} we show $S(\boldsymbol{q},t)$ at various values of $\boldsymbol{q}$. We find that the growth is most rapid along the line of modes for which
$q_{y}=0$ and the line of modes which is closest to $q_{y}\simeq q^{*}$
(the peak of the Fourier transform of the interaction potential.)
In \fig~\ref{fig:Plot-of-structure} the Monte Carlo data is averaged over at least 700 initial stripe configurations for the same parameters as the system shown in \fig~\ref{fig:OO-snapshots}.
We look at several $q_x$ modes with $q_{y}\simeq q^{*}$. There
is a steep initial increase of $S(q_x,q_y)$ and then a gradual decrease during the first Monte Carlo step, which we
interpret as stage one kinetics toward the unstable state
stripe solution. Stage one corresponds to the initial behavior of the
stripes seen in \fig~\ref{fig:OO-snapshots} followed by narrowing
of the stripes. Eventually the structure factor modes grow exponentially
before stage two kinetics breaks down at around 10\,MCS. Two distinct
exponential rates are evident. 

\begin{figure}
\includegraphics[scale=0.75]{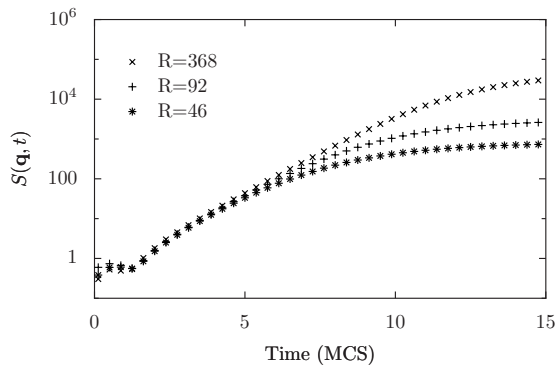}
\caption{Semi-log plot of the structure factor $S(\boldsymbol{q},t)$ at $q_{x}=0$ and $q_{y}\simeq q^{*}$
versus time in the antiferromagnetic Ising model for interaction ranges $R=46,\:92$, and $368$. The quench
parameters are the same as those in Fig.~\ref{fig:Plot-of-structure}.
We find that the time scale of stage one kinetics (about the first
MCS) is independent of $R$, and the exponential evolution
of stage two lasts longer for longer $R$.\label{fig:oo-quench-ranges}}
\end{figure}

In Fig.~\ref{fig:oo-quench-ranges} we compare the evolution of the
structure factor for systems with different interaction ranges. We
see that the time scale of stage one kinetics is independent of
the interaction range, and exponential behavior lasts longer for larger $R$, as in the CHC theory. 

\subsubsection{Understanding stage two kinetics}

To better understand the behavior of the system during stage two,
we seek an analogy to the critical quench where stage
one is entirely absent from the early unstable evolution. Thus, instead
of quenching the field, we will initially prepare our system in the
state corresponding to the unstable \el\ solution.

We verify numerically that the unstable stripe solution
$\phi_{s}(r_x,r_y)$ to the \el\ equation at a particular $T$ and $h$
corresponds to the $one-dimensional$ stable clump solution $\phi_{\rm 1D}(r_x)$
with the same temperature and external field: $\phi_{s}(r_x,r_y)=\phi_{s}(r_x)=\phi_{\rm 1D}(r_x)$.

For the stripe to clump transition we used the $h_{f}=0.8$ and $T_{f}=0.04$.
We find the stable one-dimensional solution by numerically solving \eq~\eqref{eq:relax_lang_dynamics} in one dimension at
this field and temperature. We use the corresponding unstable stripe solution $\phi_s(r_x)$ to find the matrix $A$ in \eq~\eqref{eq:matrix} which depends on $\phi_s(r_x)$ through the mobility $M(r_x)=\beta_f(1-\phi_s(r_x)^2)$. 

\begin{figure}
\includegraphics[scale=0.75]{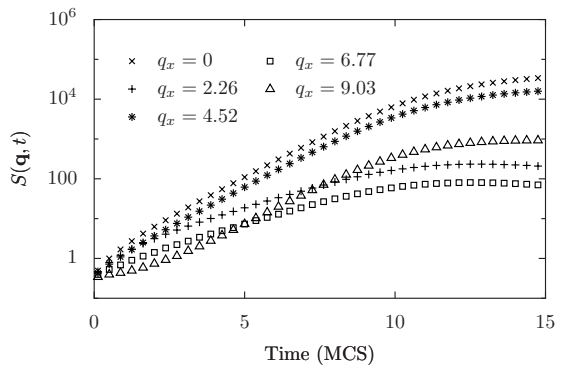}
\caption{Semi-log plot of the structure factor versus time in the antiferromagnetic Ising model for various $q_x$ values with $q_{y}=4.518$. The initial configuration is constructed
from the one-dimensional solution $\phi_{\rm 1D}$ to the \el\ equation at temperature
$T=0.04$ and field $h=0.8$. Data is averaged over about 700 independent stripe configurations. Two distinct
exponential growth rates are apparent from about 3 to 8 Monte Carlo steps.
\label{fig:Semi-log-plot-of}}
\end{figure}

\begin{figure}
\includegraphics[scale=0.75]{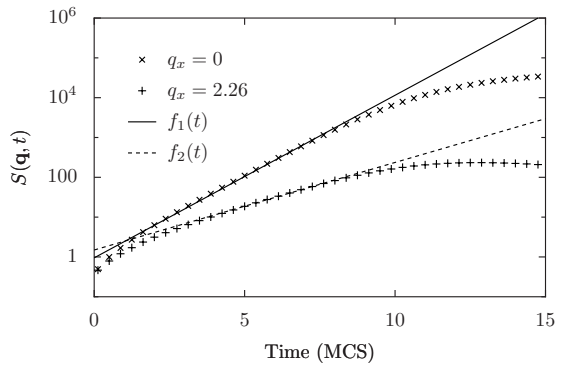}\caption{A semi-log plot of structure factor for two Fourier modes from \fig~\ref{fig:Semi-log-plot-of}
along with predictions for linear theory exponential growth, $f_{1}(t)\sim\exp(2\lambda_{1}t)$
and $f_{2}(t)\sim\exp(2\lambda_{2}t)$, where the $S$ intercept has been
fit to the data. Here $\lambda_{1}$ and $\lambda_{2}$ are the two
positive eigenvalues from our matrix calculation.\label{fig:Two-Fourier-slopes}}
\end{figure}

For the parameters considered ($h_{f}=0.8$, $T_{f}=0.04$, $L=1024$, and $L/R\simeq 2.78$), we diagonalize $A$ and find two positive eigenvalues of \eq~\eqref{eq:eigen} (Larger lattice sizes
give a larger number of positive eigenvalues. The depth of the quench also effects the number of positive eigenvalues.) We also find by examining the eigenvectors corresponding to these two 
positive eigenvalues that each Fourier mode has a non-zero contribution
from only one of the two positive eigenvalues. Thus we expect each Fourier mode to be dominated by one of the two exponential terms with a positive argument and two distinct exponential growth rates should eventually be apparent. This is consistent with the data shown in Figs.~\ref{fig:Semi-log-plot-of} and~\ref{fig:Two-Fourier-slopes}.

The one-dimensional solution is used to construct 
initial two-dimensional configurations for the Monte Carlo simulations. The length of
the one-dimensional coarse grained lattice is chosen to be the same as the
length of the two-dimensional Monte Carlo lattice.  We use the value of $\phi_{s}(r_x)$ to determine the
probability that the corresponding Ising spin is initially up: $p=(\phi_{s}(r_x)+1)/2$.
We repeat this process over all horizontal strips of the Ising lattice. We then run the Monte Carlo simulation and monitor $S(\boldsymbol{q},t)$ for various values of $\boldsymbol{q}$, averaging over independent runs. As expected, the evolution
looks similar to the evolution of the corresponding $S(\boldsymbol{q},t)$
in \fig~\ref{fig:Plot-of-structure} with about the first Monte
Carlo step removed. In \fig~\ref{fig:Two-Fourier-slopes} we have
plotted two Fourier modes from Fig.~\ref{fig:Semi-log-plot-of} along with the exponential curves from \eq~\eqref{eq:OOStructureFactor} to indicate
the growth rates predicted from our matrix calculation. 

After some initial evolution the Fourier modes eventually grow exponentially
at one of two rates, reflecting the coupling of the modes due to
the spatial variation of the initial state.

The fact that our theory predicts exponential growth rates even though
we did not consider the effect of the noise on the structure factor indicates that, as in the disorder-to-order case, the contribution
from the noise term has the same time dependence as the noiseless
part, at least at later times during stage two~\cite{noise}.

\section{summary and discussion\label{sec:summary-and-discussion}}

We have presented a generalization of the linear theory of Cahn, Hilliard, and Cook~\citep{Cahn1958,Cahn1959,Cahn1967,Cook1970}
for early stage unstable evolution. In particular, we have given a
description of the initial unstable evolution for transitions with
spatial symmetry breaking and no order parameter conservation. One of our predictions is that there are two stages of early time kinetics in general for systems with long interaction ranges. Stage one is dominated
by a symmetry preserving nonlinear dynamics and takes the system to
a symmetry constrained local minimum of the free energy on a time
scale that is independent of the interaction range. Although noise
driven fluctuations may also be growing during stage one, they do
not evolve significantly in this relatively short-lived regime. In stage
two these noise driven fluctuations break the spatial symmetry of
the initial configuration and grow for an extended period of time. 

This picture is consistent with the observed behavior of the long-range
antiferromagnetic Ising model, a system that exhibits several symmetry
breaking transitions. We find the stage two kinetics of the long-range
antiferromagnetic Ising model is characterized by long-lived exponential
growth of symmetry breaking fluctuations. For the disorder-to-order
case the symmetry breaking fluctuations are individual Fourier modes.
In contrast, the order-to-order case is complicated by the spatial
symmetry of the initial configuration, which couples the growth of
the Fourier modes. For this reason the symmetry breaking fluctuations
in the order-to-order case are not individual Fourier modes, but eigenvectors
in Fourier space. The evolution of the Fourier modes is
described by linear combinations of exponential growth and/or decay,
which after some time resembles pure exponential growth due to the
largest positive eigenvalue.

We have focused on systems without order parameter conservation.
A future publication will apply our generalized theory for order-to-order
transitions with order parameter conserving dynamics.

Our results have several consequences for experimentalists and simulators
looking to fit their data to a linear theory after an unstable quench. For
disorder-to-order transitions stage two kinetics, which is analogous
to the \chc\ theory, is expected immediately after the quench only for
systems with conserved order parameters or only in special circumstances
for systems with no conserved order parameters (as in a critical quench.)
For order-to-order transitions, exponential growth of Fourier modes
is delayed for two reasons. In stage one exponential growth
is not expected. In stage two the evolution of Fourier modes is represented
by a linear combination of exponential terms, and therefore only approximate
exponential evolution is expected after some time. Because the lifetime
of stage two kinetics is limited by the finite
interaction range, exponential Fourier mode growth may be
absent even if the linear evolution of stage two is present. 

\begin{acknowledgments}
We would like to thank Harvey Gould for his advice, suggestions, and contributions. We would also like to thank Louis Colonna-Romano, and Minghai
Li for helpful discussions. This work was funded by DOE Grant No.\
2234-5 (R. D., K. B., W. K) and NSF Grant No.\ DGE-0221680.

\bibliographystyle{apssamp.bst}

\begin{thebibliography}{37}
\bibitem{symmetry}The ferromagnetic Ising model with no external field
has an up/down spin symmetry which is broken in the final low temperature
state. This symmetry breaking should not be confused with \emph{spatial} symmetry breaking,
which is not present in the ferromagnetic Ising model.
\bibitem{analogy}In analogy to the ferromagnetic case we refer to dynamics
as conserving or not conserving the ``order parameter'' even though
$\hat{\phi}$ is not a well-defined order parameter for the antiferromagnetic
Ising model which we will consider.
\bibitem{finite} The ratio $L/R$ gives the approximate number of stripes (or clumps along
one dimension) in the system. We have tested many values of $L/R$ and found that the early stage two kinetics is unaffected by this value.  We did find that stage two lasts longer for larger values of $L/R$.  At any value of $L/R$ larger interaction ranges extend the lifetime of stage two.  We do not deal with the creation of defects
in systems with large $L/R$ because the defects do not form
until after stage two breaks down.
\bibitem{noise}That the contribution from the noise term term has the same time dependence as the noiseless term can be shown to be true in the case that $B=\textnormal{constant}$ (from \eq~\eqref{eq:generalLangevin1}),
but is complicated by further couplings for a spatially dependent
$B$.
\end{thebibliography}

\end{acknowledgments}

\end{document}